\documentclass[article]{aa}
\usepackage{graphicx}
\usepackage{natbib}
\bibpunct{(}{)}{;}{a}{}{,} 
\newcommand{\ind}[1]{_{\mathrm{#1}}}

\begin{document}

\title{SYMPA, a dedicated instrument for Jovian Seismology.\\
II. Real performance and first results}
\author{Patrick Gaulme\inst{1}
          \and
          F.X. Schmider\inst{1}
          \and Jean Gay\inst{1}
         \and
C\'edric Jacob \inst{1} \and Manuel Alvarez \inst{3} \and
   Mauricio Reyes \inst{3} \and Juan Antonio Belmonte\inst{4} \and Eric Fossat\inst{1} \and Fran\c cois
Jeanneaux\inst{1} \and Jean-Claude Valtier \inst{1} }

\offprints{P. Gaulme}

   \institute{Laboratoire FIZEAU, Universit\'e de Nice Sophia-Antipolis, CNRS-Observatoire de la C\^ote d'Azur, F-06108 Nice Cedex 2\\
              \email{Patrick.Gaulme@obspm.fr}
         \and
   Observatorio Astron\'omico Nacional, Instituto de Astronom\'ia,
 Universidad Nacional Aut\'onoma de M\'exico, Apto. Postal 877, Ensenada, B.C., M\'exico
           \and
           Instituto de Astrof\'isica de Canarias, Tenerife, Spain
           \and
           THEMIS Observatory, La Laguna, Tenerife, Spain
 }

\titlerunning{SYMPA: II. Real performance and first results}
\authorrunning {Gaulme et al.}
\abstract
{Due to its great mass and its rapid formation, Jupiter has played
a crucial role in shaping the Solar System. The knowledge of its
internal structure would strongly constrain the solar system
formation mechanism. Seismology is the most efficient way to probe
directly the internal structure of giant planets. }
{SYMPA is the first instrument dedicated to the observations of
free oscillations of Jupiter. Principles and theoretical
performance have been presented in paper I. This second paper
describes the data processing method, the real instrumental
performance and presents the first results of a Jovian observation
run, lead in 2005 at Teide Observatory.}
{SYMPA is a Fourier transform spectrometer which works at fixed
optical path difference. It produces Doppler shift maps of the
observed object. Velocity amplitude of Jupiter's oscillations is
expected below 60 cm s$^{-1}$. }
{Despite light technical defects, the instrument demonstrated to
work correctly, being limited only by photon noise, after a
careful analysis. A noise level of about 12 cm s$^{-1}$ has been
reached on a 10-night observation run, with 21\% duty cycle, which
is 5 time better than previous similar observations. However, no
signal from Jupiter is clearly highlighted. }
   {}

\maketitle

\section{Introduction}

Due to its great mass and its rapid formation, Jupiter has played
a crucial role in shaping the Solar System. Two scenarios are
generally proposed for the formation of giant planets: the
nucleated instability (Safronov \& Ruskol 1982) and the
gravitational instability models (Cameron 1978 and Mayer et al.
2002). An efficient constraint on the formation scenario would be
given by measuring the total amount of heavy elements inside
Jupiter and the size of the planetary core. Moreover, the
knowledge of Jupiter's internal structure would constrain the high
pressure hydrogen equation of state, which is still inaccurate,
and would particularly solve the question of the nature of the
metallic-molecular phase transition (e.g. Guillot et al. 2004).
Gudkova \& Zharkov (1999) showed that the observation of
oscillation modes up to degree $\ell=25$ would strongly constrain
Jupiter's internal structure by exploring both the hydrogen
plasma-phase transition and the supposed core level.

Attempts to observe Jovian oscillations have been brought since
the mid 1980's, thanks to different techniques. On one hand,
Deming et al. (1989) have looked for oscillation signature in
thermal infrared. Unfortunately, their infrared detectors, not
enough sensitive, did not detect any signal. On the other hand,
oscillations were sought in velocity measurements, obtained by
Doppler spectrometry. Schmider et al. (1991, hereafter S91) used
sodium cell spectrometer and Mosser et al. (1993, 2000, hereafter
M93 and M00) the Fourier transform spectrometer FTS (at CFHT,
Hawaii) at fixed optical path difference. An excess of power has
been brought out in the spectrum at frequency range [0.8-2] mHz,
as well as the large separation of oscillation $p$-modes around
140 $\mu$Hz. Nevertheless, the oscillation modes have never been
individually identified, hindering any constraint about the
internal structure.

SYMPA is an instrument dedicated to Jovian oscillations, which
concept and performance have been described in paper I (Schmider
et al. 2007). For the first time, a specific instrument dedicated
to Jovian oscillations was developed, including imaging
capability. Indeed, full disc observations do not permit to
distinguish modes of degree higher than 3. Moreover, the
broadening of the solar lines, due to the fast rotation of
Jupiter, reduces drastically the sensitivity of such measurements.
The instrument, a Fourier tachometer, is composed of a
Mach-Zehnder interferometer which produces four images of the
planet, in the visible range corresponding to three Mg solar
absorption lines at $517$ nm. The combination of the four images,
in phase quadrature, allows us to retrieve the phase of the
incident light, which is related to the Doppler shift generated by
the oscillations.

Two instruments were built at Laboratoire Fizeau (Nice
University). Three campaigns were lead simultaneously on two
sites: in 2003, at San Pedro Martir (Mexico) and Calern (France)
observatories; in 2004 and 2005, at San Pedro Martir and Teide
(Canaries) observatories. 2003 campaign was mainly dedicated to
technical commissioning. In Canaries, bad weather conditions have
strongly limited the efficiency of the 2003 and 2004 campaigns,
whereas the 2005 campaign benefited of better conditions.

In this paper we present the data processing, the real performance
and the first results, obtained during the 2005 run at Teide
observatory. The processing of the San Pedro Martir data and the
combined analysis of both observing 2005 network campaign will be
considered in a future work. After a short presentation of the
observing conditions (Sect. \ref{campagne}), Sect. \ref{process}
presents the process of the data analysis. We expose in Sect.
\ref{calibration} all the steps for an accurate calibration of the
data. Sect. \ref{singleimage} is devoted to the further of the
data reduction, for the measurement of velocity maps. The analysis
of the time series get during the 2005 run at Tenerife is exposed
in Sect. \ref{single_nite}. Section \ref{conclu} is devoted to
conclusions and prospectives.

\section{Observations}
\label{campagne}

Observations were conducted at Teide observatory (Canaries
islands), with the 1.52-m Carlos Sanchez telescope, between march
31st and April 10th 2005. As it has been detailed in paper I, four
images of Jupiter come out from the instrument. Optical parameters
inside SYMPA's box are arranged such as the four 1.3-arcmin
fields, cover 128 pixels on the receptor (a DTA CCD, $1024\times
256$ pixels). During the run, Jupiter at opposition presented a
diameter of 48 arcsec, corresponding to 69 pixels on the CCD
camera.

\begin{table*}
\centering
 \caption{2005 run at Teide Observatory}
\begin{tabular}{|c|c|c|c|c|c|c|}\hline
 Starting date & Ending date &Duration & Mean sampling& Number of & Mean Intensity & Standard
 deviation\\
m d h m s (UT)  &m d h m s (UT) & h m s & s &acquisitions & photons image$^{-1}$ & photons image$^{-1}$\\
 \hline
Apr-02, 23:48:14& Apr-03, 07:03:40 & 7:09:35& 5.45 &4727 & 1.89 $10^9$& 0.67 $10^8$  \\
Apr-03, 23:30:14& Apr-04, 07:02:29 & 7:31:59& 7.13 &3792 & 1.79 $10^9$ & 5.27 $10^8$\\
Apr-04, 23:16:12& Apr-05, 06:58:17 & 7:39:30& 6.75 & 4080& 2.50 $10^9$& 0.76 $10^8$\\
Apr-05, 23:25:29& Apr-06, 06:50:23 &7:22:12&6.68 & 3971& 2.44 $10^9$& 3.80 $10^8$\\
Apr-09, 23:20:17& Apr-10, 06:31:59 & 6:37:48&6.28 & 3799 & 2.06 $10^9$& 3.62 $10^8$\\
Apr-10, 22:51:11& Apr-11, 06:33:11 & 7:41:48&6.33 & 4376&2.38 $10^9$&1.14 $10^8$\\
Apr-11, 21:09:35& Apr-12, 06:30:17 & 7:20:42& 6.83&3872 &2.36 $10^9$&0.97 $10^8$\\
 \hline
\end{tabular}
\label{tab:campagne}
\end{table*}

\begin{figure}
\includegraphics[width=8.4cm,height=4.2cm]{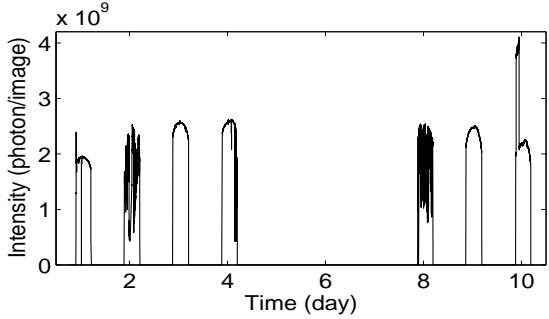}
\caption{Total intensity of each image. Nights 2 and 8 and the end
of night 4 present a strongly and rapidly variable amount of
photons, because of cloudy conditions.}
\label{fig:SNR}
\end{figure}

\begin{figure}
\label{window}
\includegraphics[width=8.4cm,height=4.2cm]{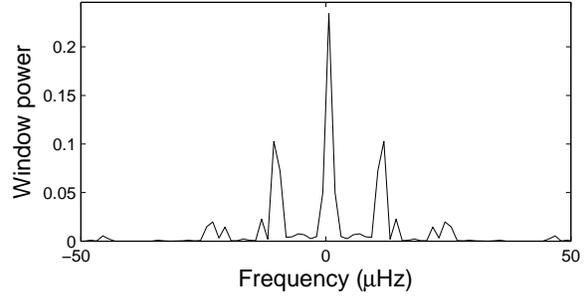}
\caption{Spectral window of the observation campaign. The
amplitude has been normalized to the spike's power. Only 23\% of
the total energy remains in the central peak. }
\end{figure}

Seven nights over ten days benefited of good weather conditions,
yielding to a 21\% duty cycle. Data quality was almost constant
from one night to another, excepted for night 2 and 8, where
clouds have reduced the incident flux. Observation conditions are
summarized in Fig. \ref{fig:SNR} and Table \ref{tab:campagne}. The
window function associated to the whole campaign was very strong,
since we consider only Canaries data. Therefore, in the power
spectrum, the amplitude of a single spike has been divided by a
factor 4, since its power gets diluted in high side lobes (Fig.
\ref{window}). The total flux was expected to be about $2.4\ 10^9$
photons per 6-s exposure (paper I). However, its mean value along
the run is about $2.2\ 10^9$ photons per exposure (Fig.
\ref{fig:SNR}). This discrepancy with the estimated flux
introduces a factor $0.95$ in the ratio to noise value. In paper
I, a noise level of about 4 cm s$^{-1}$ was expected for a
16-nights observation campaign, with 50\% duty cycle. Instead of
such a performance, by considering only Canaries data, the noise
level is therefore expected at 10 cm s$^{-1}$.

\section{Data processing strategy}
\label{process}
\subsection{Four interferograms in phase quadrature}

\begin{figure}
\includegraphics[width=8.4cm]{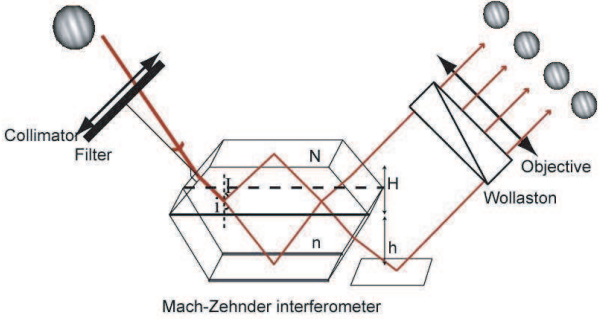}
\caption{Schematic view of the SYMPA instrument. The incident
light coming from the 1.5-m telescope, passes through a 120-mm
collimator and the 5 nm bandwidth interference filter. The optical
path difference $\Delta$ occurs inside the Mach-Zehnder prism; it
is a function of the heights $H$ and $h$, refraction index $N$ and
$n$ and incidence angles into the prism $I$ and $i$ : $\Delta = 2
(H N \cos I - h n \cos i$. The wollaston polarizer separates each
output from the interferometric device into two separated beams.
In total, the instrument produces on the camera four images of the
same field, separated by $\pi/2$ in phase (see paper I).}
\label{fig_SYMPA_3D}
\end{figure}

SYMPA's instrumental principles are fully explained in paper I and
are summarized in Fig. (\ref{fig_SYMPA_3D}). The four output beams
can be described in the detector coordinates $(x,y)$ by the
following approximations :
\begin{eqnarray}
I_1(x,y) &=& \frac{I_0(x,y)}{4}\ \left[1\ -\ \gamma\cos \phi(x,y)\right] \label{I1}\\
I_2(x,y) &=& \frac{I_0(x,y)}{4}\ \left[1\ -\ \gamma\sin \phi(x,y)\right] \label{I2}\\
I_3(x,y) &=& \frac{I_0(x,y)}{4}\ \left[1\ +\ \gamma\cos \phi(x,y)\right] \label{I3}\\
I_4(x,y) &=& \frac{I_0(x,y)}{4}\ \left[1\ +\ \gamma\sin
\phi(x,y)\right] \label{I4}
\end{eqnarray}
where $I_0$ is the continuum component of the incident light, that
is to say the Jovian figure, $\gamma$ the fringe contrast and
$\phi(x,y)$ the incident wave phase map :
\begin{equation}
\label{phase}
\phi(x,y) = 2\pi \sigma_0 \Delta(x,y) \left(1 +
\frac{v\ind{D}}{c}\right)\label{phase}
\end{equation}
where $\sigma_0$ is the central wavenumber of the input filter,
$\Delta(x,y)$ is the optical path difference (OPD) and $v\ind{D}$
and $c$ are the Doppler and the light velocities. The Doppler
shift of the solar Mg lines comes from the combination of the
relative motion of Jupiter to the Sun $v\ind{J/S}$, relative
motion of the observer to Jupiter $v\ind{E/J}+ v\ind{E,rot}$
(distance between the two planets and Earth's rotation), Jupiter's
rotation $v\ind{J,rot}$ and, finally, the oscillations
$v\ind{osc}$. In the following, we write the Doppler velocity as
the sum of :
\begin{equation}
v\ind{D} = 2\left( v\ind{J/S} + v\ind{E/J} + v\ind{E,rot} +
v\ind{J,rot} + v\ind{osc}\right) \label{dopplervelocity}
\end{equation}
The factor 2 is due to the fact that the Doppler effect gets
doubled after reflection on Jupiter's atmosphere. The orders of
magnitude of these different terms are presented in table
\ref{vitesses}.

From Eqt. \ref{phase}, the phase map appears to be the sum of two
contributions : a ``motionless'' term, $2\pi \sigma_0
\Delta(x,y)$, and a velocity term, $4\pi \sigma_0 \Delta(x,y)
v\ind{D}/c$. The art of extracting the oscillation signal resides
in the ability of eliminating step by step the motionless fringes
and the ``spurious'' velocity fields $v\ind{J/S}$, $v\ind{E/J}$,
$v\ind{E,rot}$ and $v\ind{J,rot}$. All the steps of the schematic
process presented below are detailed in Sect. 4 and 5.

\begin{table}[h]
\begin{center}
\caption{Orders of magnitude of different Doppler shifts during
the 2005 run at Teide Observatory.}
\begin{tabular}{|l | c |}\hline
 & Velocity (m s$^{-1}$) \\
\hline\hline
Jupiter-Sun      & 0.4   \\
Jupiter-Earth    & 3142   \\
Earth rotation   & $[-409, 409]$  \\
Jupiter rotation & $[-12570, 12570]$   \\
Oscillations & $<0.6$ \\
\hline
\end{tabular}
\label{vitesses} \end{center}
\end{table}

\subsection{Extracting the oscillation signal}

Let us consider a quadruplet of interfering images on the
detector's field. The differences between the two couples of
images, which are in phase opposition, allow to cancel the
continuous component of the interferograms $I_{0}$, in order to
keep the interfering patterns. We write $U$ and $V$ the normalized
interfering patterns :
\begin{eqnarray}
U &=& \frac{I_1 - I_3}{I_1+I_3}\ \propto\ \gamma \cos \phi \label{U}\\
V &=& \frac{I_2 - I_4}{I_2+I_4}\ \propto\ \gamma \sin \phi
\label{V}
\end{eqnarray}
where we let go of $(x,y)$ dependence in order to simplify
notations. The incident wave phase is retrieved by taking the
argument of the complex interferogram:
\begin{equation}
Z = U + \imath V\ \propto\ \gamma\ e^{\imath \phi}
\end{equation}
The data processing expands in three main steps: correction of the
motionless fringes, elimination of Jupiter rotation and
elimination of the relative motion of the observer to the target
and of the target to the Sun. Mathematically, it lies in creating
successively four complex interferograms, associated to each step
of the data processing: $Z_0$ for motionless fringes,
$Z\ind{J,rot}$ and $Z\ind{E,rot}$ for Jupiter's and Earth's
rotation, $Z\ind{E/J}$ for the relative motion of the Earth to
Jupiter and $Z\ind{J/S}$ for Jupiter's motion with respect to the
Sun. Then, the rough Jupiter complex interferogram $Z\ind{jup}$
comes deconvoluted from additive signals:
\begin{eqnarray}
Z\ind{J,flat} &=& Z\ind{jup}\times Z_0^\ast\times
Z\ind{J,rot}^\ast Z\ind{E,rot}^\ast
\times Z\ind{E/J}^\ast\times Z\ind{J/S}^\ast \label{Z_flat_1}\\
 & \propto & \exp\left(4\pi \sigma_0 \Delta
\frac{v\ind{osc}}{c}\right) \label{Z_flat_2}
\end{eqnarray}
where the asterisk indicates the complex conjugation. The
resulting interferogram is so-called $Z\ind{J,flat}$ because of
the appearance of its argument, i.e. the velocity map. Indeed,
since oscillation amplitude is expected to be lower than 0.6 m
s$^{-1}$, it is absolutely impossible to see directly oscillation
modes in a single phase map, where mean noise level is expected to
be about 900 m s$^{-1}$ per pixel (Paper I). Oscillations might be
picked out only in the spectrum of long time series.

\begin{figure}
\includegraphics[width=8.4cm]{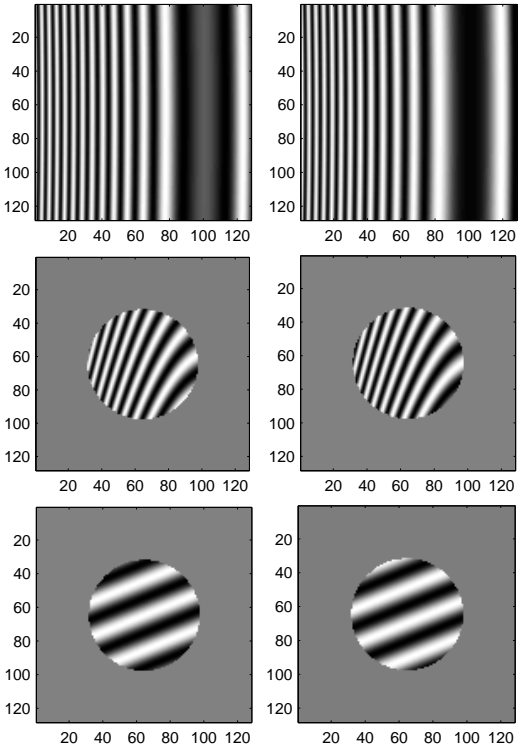}
\caption{Simulation of interferograms along the data processing
chain. Top: motionless interferograms $\left(U_0, V_0\right)$.
Middle: Jovian interferogram $\left(U\ind{jup}, V\ind{jup}\right)$
when Jupiter is inclined of $-72^\circ$ with respect to vertical.
The interference pattern is mainly due to the coupling between
motionless fringes and Jovian rotation (Eqt.
\ref{dopplervelocity}). Bottom:
 Jovian interferograms deconvoluted from motionless contribution,
 $\left(U\ind{rot}, V\ind{rot}\right)$, i.e. fringe pattern associated to Jovian rotation.
 Note that fringes associated
 to solid rotation present velocity iso-values parallel to the rotation
 axis. Differential wind profile has not been introduced.}
 \label{methode}
\end{figure}

Note that Doppler shifts due both to Earth relative motion to
Jupiter and Jupiter relative motion to the Sun are uniform across
the field. On the contrary, the Jovian Doppler component is very
sensitive to Jovian rotation since its value varie from $-12.57$
km s$^{-1}$ to $12.57$ km s$^{-1}$ from east to west on Jovian
equator. In terms of velocity, it implies that the error in
positioning the phaser $Z\ind{J,rot}$ must be smaller than the
photon noise. Therefore, it supposes to know Jupiter's position
better than 1/20 th of pixel. Figure \ref{methode} presents
simulations of the motionless interferograms, Jovian
interferograms and Jovian rotation interfering pattern.

\section{Data calibration}
\label{calibration}
\subsection{Pre-processing operations}

Pre-processing consists in cleaning each quadruplet of Jovian
image $(I_1,I_2,I_3,I_4)$ in order to create couples of
interferograms $(U,V)$. This implies three main operations. First,
the camera dark current contribution has to be subtracted, by
using offset images. Second, the inhomogeneities of the single
pixel responses to light intensity (photon/electron gain) have to
be compensated by dividing each image by a flat field image.
Third, the construction of the Jupiter phase map, using the
argument of the complex image created by the difference of two
couples of images, requires that all the four images overlap one
to each other. If the two first points are easy to realize, the
last one is pretty delicate because of the required accuracy of
about 1/20 pixel.

\begin{figure}
\includegraphics[width=4.9cm]{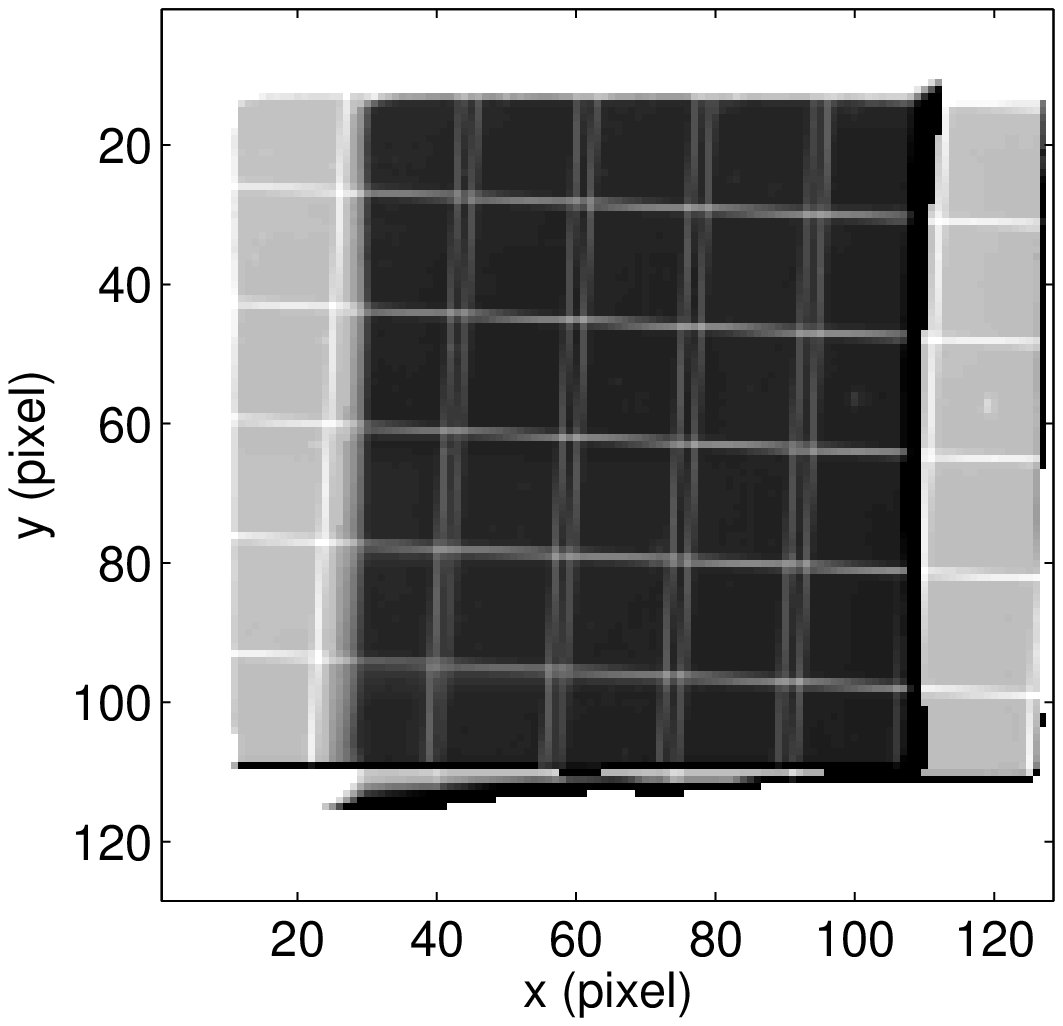}
\hspace{-1cm}
\includegraphics[width=4.9cm]{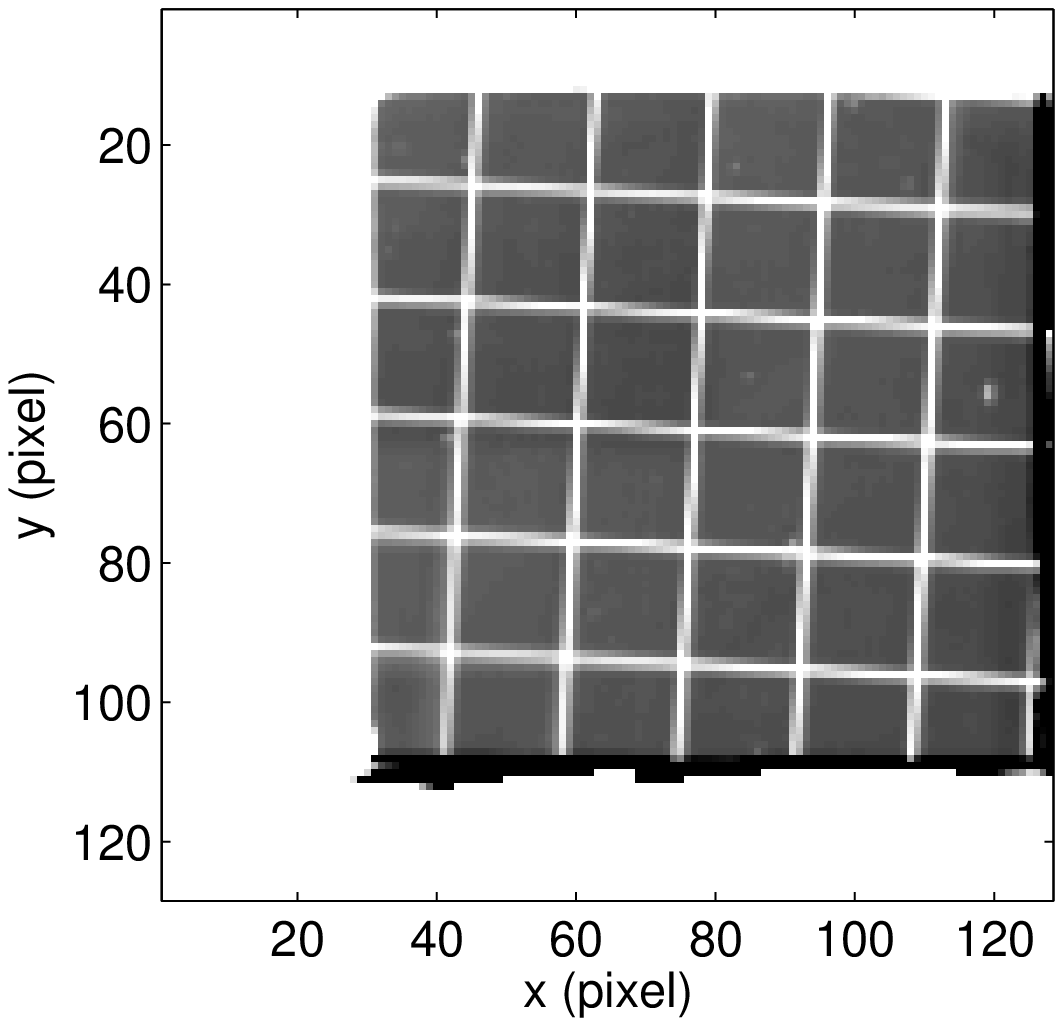}
\caption{Overlapping of grids 1 and 3, before then after
distortion rectification. The position of the horizontal and
vertical lines are determined by fitting the zeros of the grid
image derivatives, along both directions $x$ and $y$. Second order
polynomials are sufficient, since rotation and barrel distortion
do not require higher orders. Then, each intersection point
coordinates are calculated by solving numerically the four order
equation coming from the combination of the vertical and the
horizontal line fits. The mean distance between intersections
points is equal to 5.32 \% pixel$^{-1}$}
\label{grilles_avant_apres}
\end{figure}

As for every optical system including lenses and prisms, field
distortion is unavoidable. Although Jupiter was positioned as
close as possible to the optical axis, its large diameter involves
differential distortions between the four images, making the
overlapping impossible. The whole distortion effect is supposed to
be composed of only translation, rotation and barrel distortion.
This problem has been anticipated by putting a regular grid,
engraved on a glass slice, at the instrument focus. The grid
intersection positions are used to characterize the distortion
(Fig. \ref{grilles_avant_apres}).

Let us consider one image among the quadruplet $I_i(x,y)$, $i \in
[1,4]$, where $(x,y)$ are the detector coordinates (CCD pixels).
Because of optical distortion, the image value on the $(x_k,y_k)$
point, $k \in [1,128]$, actually corresponds to the $(x_k',y_k')$
point. The repositioning algorithm must produce for each image
$I_i$ an image $I_i'$, defined in the detector's coordinates, as:
\begin{equation}
I_i'(x,y)\ =\ I_i(x',y')
\end{equation}
The distorted coordinates $(x',y')$ are related to the regular
detector coordinates by the relations:
\begin{equation}
x'\ =\ x + f(x,y)\qquad\mbox{and}\qquad y'\ =\ y + g(x,y)
\end{equation}
where $f$ and $g$ are polynomials expressed as $\sum_k C_k x^j
y^{k-j}$, where $k$ is the polynomial order and $j \le k$. The
polynomial coefficients $C_k$ are obtained by minimizing the
difference between the intersection point coordinates of the grid
associated to the considered image $I_i$ and the intersection
point coordinates of the regular grid. The knowledge of the set of
distorted coordinates $(x',y')$ is used to build the new rectified
image $I_i'(x,y)$, by interpolation of $I_i(x',y')$ upon the
regular detector coordinates.

Interpolation is realized with cubic method. In Fig.
\ref{grilles_avant_apres} we present a couple of 2-grid images
before and after repositioning. Note that the translation
repositioning represents the zero order of the transformation. A
way to evaluate the accuracy of the repositioning process is to
apply it to the previously repositioned grids; they should overlap
(Fig. \ref{grilles_avant_apres}, right). The mean distance between
the repositioned intersection points from one grid to another
belongs to the range $[1/20, 1/15]$ pixel, which almost fulfills
the required accuracy.

In order to respect the conservation of the flux, correcting
operations have to be processed in the following order: first,
subtraction of the offset to the image and to to the flat-field,
then rectification of the offset-corrected image and flat-field.
\begin{equation}
I\ind{pre-processed}\ =\ \frac{\left[I(x,y,T) -
O(x,y,T)\right]\ind{rectified}}{\left[F(x,y,T) -
O(x,y,T)\right]\ind{rectified}}
\end{equation}
where $I$ indicates the considered image (e.g. Jupiter), $O$ the
offset image and $F$ the flat-field image. T stands for the
temperature of the camera. Note that assessing the four output
intensities are not strictly equal, no photometric balance has to
be performed since it is implicitly done by dividing each image by
the flat field.

\subsection{Motionless fringes calibration}

The next step of data processing is the deconvolution of Jovian
complex interferograms $Z\ind{jup}$ from motionless interferogram
$Z_0$. Therefore, we have to characterize the motionless phase
term $2\pi \sigma_0 \Delta(x,y)$ (Eqt. \ref{phase}). Furthermore,
as it has been reported in paper I, the four output beams are not
in perfect phase quadrature: the discrepancies between actual
measurements and theoretical expectation are about $\varepsilon =
28^\circ$. This shift compared to quadrature has to be quantified
precisely across the whole field $(x,y)$.

The knowledge of the optical parameters of both telescope and
instrument permits to describe the motionless fringes (cf
simulations on Fig. \ref{methode}), but not their imperfections.
Strong constraints on the motionless fringes come from solar light
scattered by the telescope dome. Indeed, excluding the spectral
Doppler shift, it presents the same spectrum as Jupiter, since
517-nm magnesium lines are solar reflected lines. Moreover such a
process enables to enlighten all the detector surface. Actually,
the only difference between motionless fringes and ``sky'' fringes
is the uniform velocity field, introduced mainly by Earth
rotation, much less by Sun-Earth distance variation and by
undesired thermal effects (see Sect. \ref{single_nite}). These
parameters are taken into account thanks to the IMCCE ephemeride
data base (www.imcce.fr).

6-hour long ``sky'' shots were taken on April 2nd. In order to
increase the signal to noise ratio of sky interferograms, images
are averaged over one minute intervals, which corresponds to three
images. Figure ({\ref{cielsbruts}) presents one of the 330 couples
of 1-minute exposure interferograms. Fringes are well defined and
fringe contrast ($\simeq$ 0.4\%) is smaller than expected (0.8
\%). Unfortunately, it appears that the map of both interferograms
are not plane, but show a parabolic curvature. The origin of such
a residual signal is explained by the fact that sky shots are
taken during the day: some direct light enters into the
instrument, and does not follow the correct optical path.

\begin{figure}
\includegraphics[width=8.4cm]{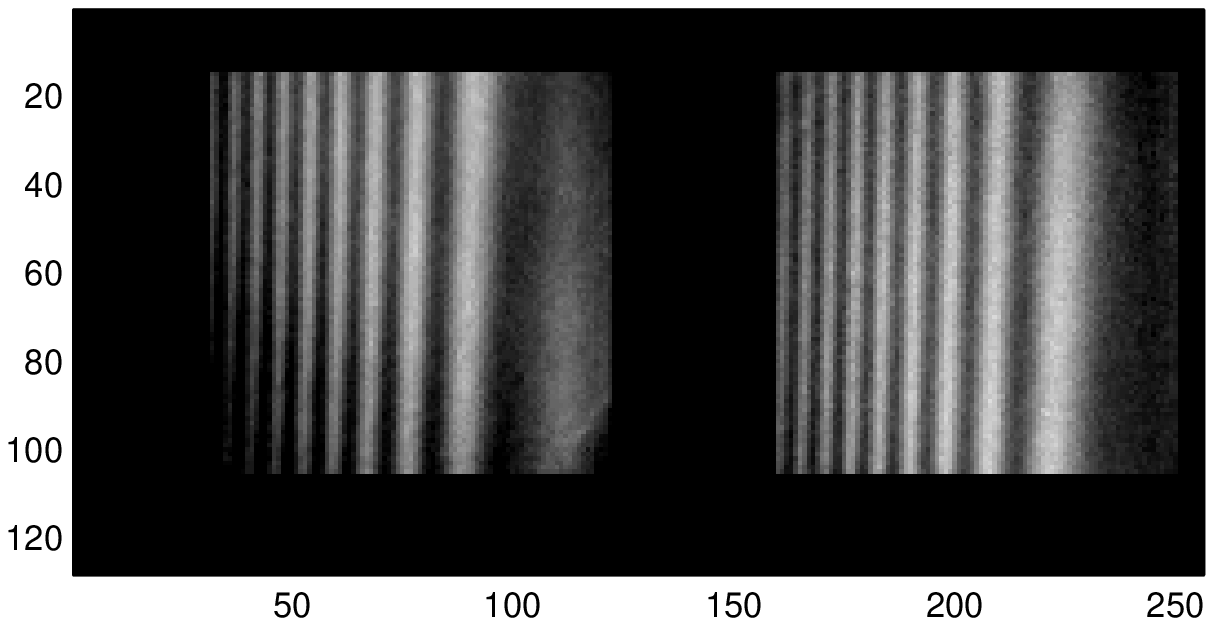}
\includegraphics[width=8.4cm]{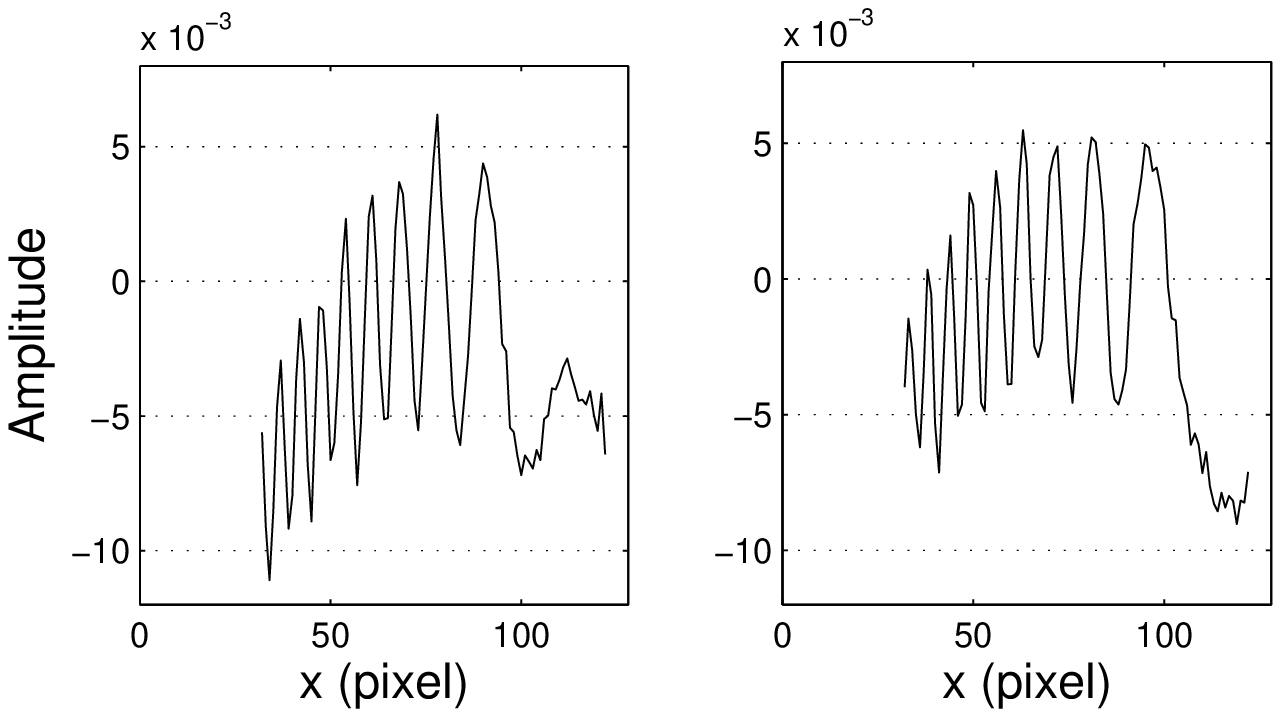}
\caption{Top: interference pattern $(U,V)$ obtained with solar
light scattered by the telescope dome, over one minute (sum of 3
images). Bottom: cut along $x$-axis of $U$ and $V$, at $y = 55$
pixel. Note that both interferograms present a varying contrast
between 0.3 and 0.5 \% and a significant unflatness. Moreover,
they are not centered around 0, that makes the phase $\phi=$
arg$(U+iV)$ impossible to recover.} \label{cielsbruts}
\end{figure}
\begin{figure}
\includegraphics[width=8.4cm]{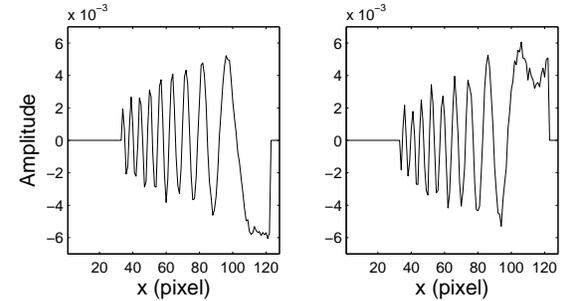}
\caption{Cut of interference patterns $(\Delta U,\Delta V)$ along
$x$-axis. After correction of the low spatial frequency term,
fringe contrast still varies strongly, from 0.2 \% to 0.6\%.}
\label{deltaUdeltaVcut}
\end{figure}

A method has been developed to get round this unflatness. Let us
take two couples of interferograms $(U,V)$ and $(U',V')$, taken at
two different dates. The phase difference $\delta \phi$ between
them is due to terrestrial motion and temperature variation. Their
expressions are given by :
\begin{eqnarray}
U &=& U_0 + \gamma\ind{U} \cos(\phi)  \ \qquad U' = U_0 + \gamma\ind{U}\cos(\phi + \delta \phi)\\
V &=& V_0 + \gamma\ind{V} \sin(\phi + \varepsilon) \ \ \ V' = V_0
+ \gamma\ind{V}\sin(\phi + \delta \phi + \varepsilon)
\end{eqnarray}
where $(U_0,V_0)$ describe the unflatness, and
$(\gamma\ind{U},\gamma\ind{V})$ the fringe contrasts. The
subtraction of these signals eliminates the unflatness terms and
brings to the two interfering patterns :
\begin{eqnarray}
\Delta U &=& U' - U = -\gamma\ind{U} \left[2 \sin\left(\frac{\delta \phi}{2}\right)\right] \sin(\varphi)\\
\Delta V &=& V' - V = \ \ \gamma\ind{V} \left[2
\sin\left(\frac{\delta \phi}{2}\right)\right]
\cos(\varphi+\varepsilon)
\end{eqnarray}
where $\varphi$ is the mean phase $\varphi = \phi + \delta
\phi/2$. Figure \ref{deltaUdeltaVcut} shows cuts of $(\Delta
U,\Delta V)$ along the $x$-axis; the unflatness problem has been
corrected. The fringe contrast presents optimized variations
across both fields, from 0.2\% to 0.6\%.

\begin{figure}
\includegraphics[width=3.7cm]{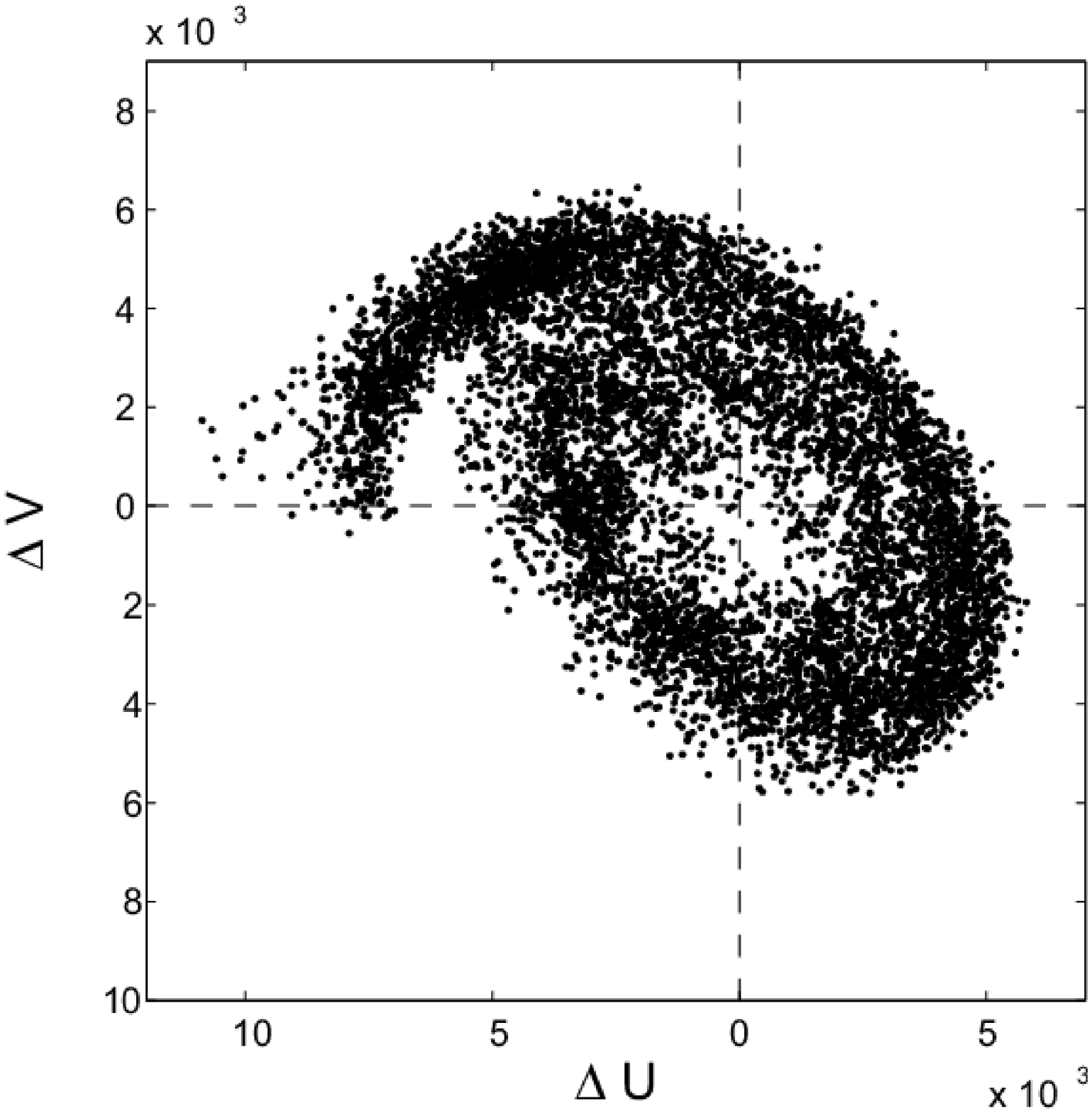}
\includegraphics[width=3.8cm]{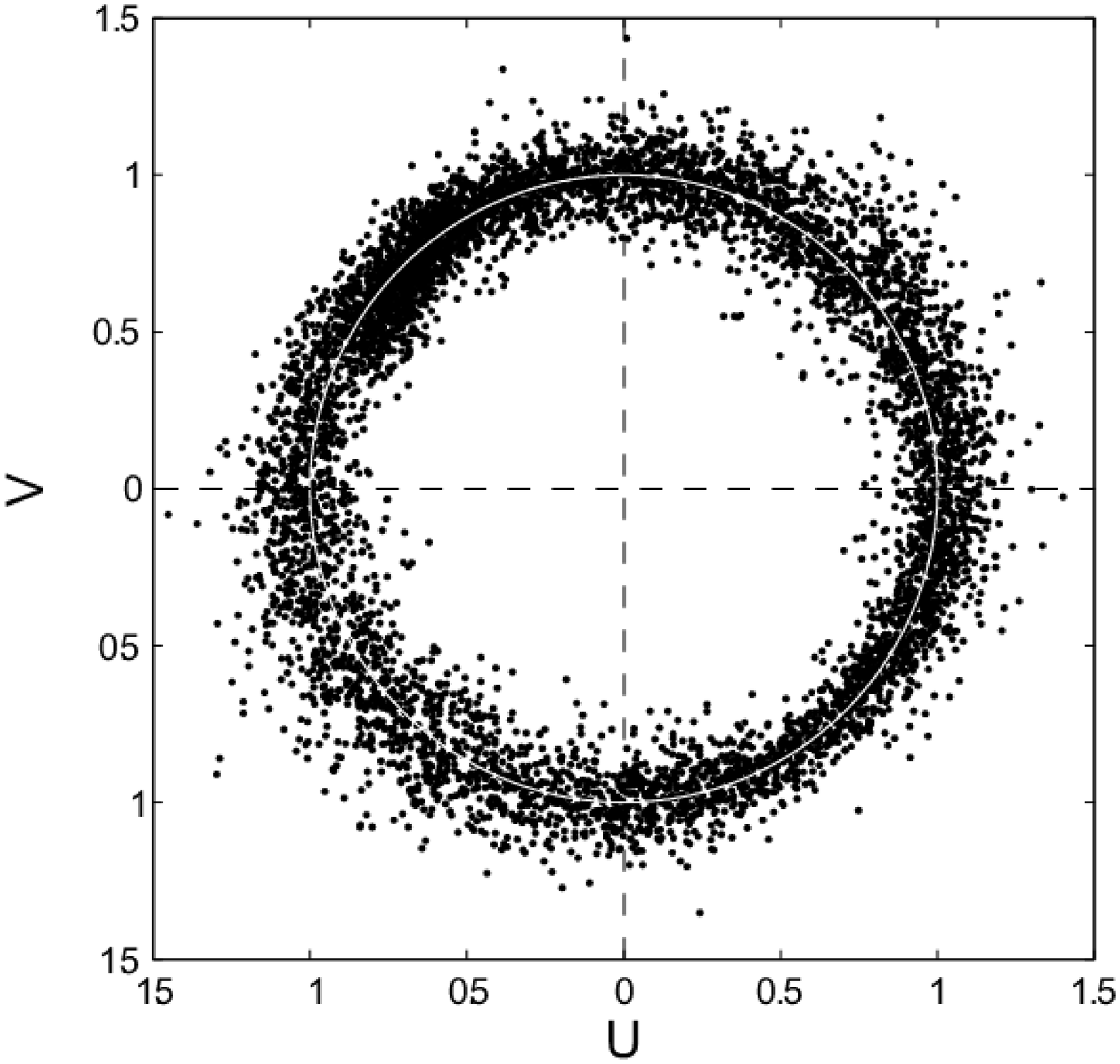}
\caption{Left: plot of the interference pattern $\Delta U$ as a
function of the other $\Delta V$. The spiral structure comes from
the fact that both fringe contrast and ellipticity vary across the
field. Right: plot of the interference pattern $\Delta U'$ as a
function of the other $\Delta V'$. Points are distributed along a
circle (solid line, white), which indicate that the two
interferograms are in phase quadrature. In terms of phase
accuracy, the standard deviation of these points is $5.3^\circ$. }
\label{ciel_complexe}
\end{figure}
The motionless complex interferograms $Z_0 = U_0 + \imath V_0$ is
obtained by fitting both quadrature shift and phase on the couple
$(\Delta U,\Delta V)$. First, the plot of one of the interference
pattern as a function of the other highlights the phase quadrature
imperfections (Fig. \ref{ciel_complexe}). In case of perfect
quadrature and uniform contrast in both interferograms, points
would be distributed around a circle centered on 0. As a result,
points are distributed along an ellipse, whose radius varies
strongly on detector field. The phase shift $\varepsilon$ with
respect to quadrature
 is related to the ellipse parameters $(A,B)$ by:
\begin{equation}\label{ecartquad}
\varepsilon = \arcsin \left(\frac{A^2 - B^2}{A^2 + B^2}\right)
\end{equation}
Since ellipticity and amplitude vary across the field (Fig.
\ref{ciel_complexe}, left), a further correction has to be
applied. Therefore, the field $(x,y)$ is divided in 10-pixel large
squares, in which all these parameters are considered as uniform.
Ellipse axis are estimated by least square fitting. The resulting
parameters estimate $(A,B)$, obtained for each sub-region, are
interpolated for each pixel, by fitting their values by a 4th
order polynomial. Thereafter, thanks to $A$, $B$ and
$\varepsilon$, the amplitude of both interference patterns
$(\Delta U,\Delta V)$ is normalized to 1 and the phase shift is
set to $90^\circ$, by the operation:
\begin{eqnarray}
\Delta U' = \frac{\Delta U\ind{norm} \cos (\varepsilon/2) - \Delta V\ind{norm}  \sin(\varepsilon/2)}{\cos(\varepsilon/2)} \label{circu_U}\\
\Delta V' = \frac{\Delta V\ind{norm} \cos (\varepsilon/2) - \Delta
U\ind{norm} \sin(\varepsilon/2)}{\cos(\varepsilon/2)}
\label{circu_V}
\end{eqnarray}
where the subscript ``norm" indicates that amplitudes have been
normalized to 1. These new variables are now in the required
quadrature. Figure \ref{ciel_complexe} shows the plot of the
interference pattern $U$ as a function of the other $V$; phase
quadrature is reached. At last, the motionless phase
$\phi\ind{instru}$ is obtained by fitting with a 4th order
polynomial the argument of the complex interferogram $Z\ind{sky} =
\Delta U' + \imath \Delta V'$. The resulting phase standard
deviation is about $5.3^\circ$.

The motionless interferogram used in the following to process
Jovian data is simply built as:
\begin{equation}\label{eq_Zinstru}
Z_0 = e^{\displaystyle \imath \phi\ind{instru}}
\end{equation}
Indeed, since only the phase term matters, the fringe contrast is
set equal to 1 in the whole field.

\section{From four Jupiter images to a velocity map }
\label{singleimage}

Processing the data lies in turning each quadruplet of Jovian
images into a calibrated radial velocity map. The first step,
developed in the previous section, gives clean fringes on both
interference patterns $(U\ind{jup},V\ind{jup})$, in order to
create the complex Jovian interferogram $Z\ind{jup}$. Second,
motionless phase is deconvoluted to $Z\ind{jup}$ with the help of
$Z_0$ (Eqt. \ref{eq_Zinstru}). Third, Jovian rotation and other
uniform velocity drifts have to be deconvoluted, in order to
extract the velocity map.

\subsection{Cleaning Jovian fringes}
\label{clean}

Let us consider a quadruplet of 4 pre-processed Jovian images.
Since they have been corrected from flat field and optical
distortions, each couple of images $(I_1,I_3)$ and $(I_2,I_4)$
present the same intensity level and shall overlap (Fig.
\ref{jupiter_quadruplet}). Then, the two interfering patterns
$(U\ind{jup},V\ind{jup})$ are created following Eqts. \ref{U} and
\ref{V}. Moreover, they are set in phase quadrature following
relation (\ref{circu_U}) and (\ref{circu_V}). As it can be seen in
Figs. (\ref{jupiter_U_V_brut}) interference fringes appear, but,
as for the ``sky" interferograms, the fringes do not oscillate
around 0, but around a distorted surface. These features are
photometric residues, which have not disappeared with operations
(\ref{U}) and (\ref{V}). However, fringes have to be repositioned
around a flat surface, centered around 0, in order to follow the
phase of the Doppler signal.
\begin{figure}
\includegraphics[width=9cm]{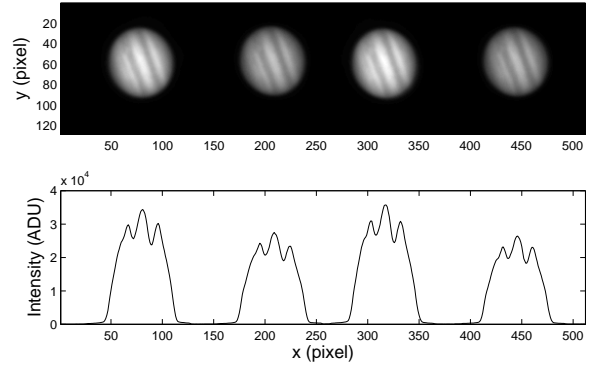}
\caption{Top: quadruplet of Jupiter images $(I_1,I_2,I_3,I_4)$,
taken during night of April 2nd 2005 at Teide observatory. Bottom:
cut of the top images. The instrument exhibits a global
polarization effect, resulting in a better transmission of one
channel with respect to the other. } \label{jupiter_quadruplet}
\end{figure}

The best way to separate the photometric noise and spurious signal
resides in filtering in the spatial frequency domain. The two
dimension fast Fourier transform (FFT) is applied to the complex
interferogram $Z\ind{jup}$. In order to avoid spectral leakage due
to the finite size of the image, we apply an oversampling on the
data $Z\ind{jup}$, by a factor 2, before passing to the Fourier
domain. Besides, Jupiter's boundary is apodized with a $\cos^3$
function, to avoid Airy-like rebounds due to Fourier transform. In
Fig. \ref{jupiter_filtrage_fourier}, we present the Fourier
transform modulus. The central region contains the low frequency
information (mean value, slow distortions). The horizontal line
corresponds to the interference pattern; the horizontal structure
comes from the fact that optical path difference varies
essentially with $x$-coordinates. The inclined alignment of stains
constitutes the photometric effect. Indeed, it corresponds to the
spatial frequencies along the rotation axis of Jupiter, which is
inclined of about $-72^\circ$ in the detector's field. These
features are actually remnants of cloud zones and belts.

Wiener filtering stipulates that signal can be filtered out from
noise in the Fourier domain if they were clearly differentiable.
As it can be seen in Fig. \ref{jupiter_filtrage_fourier}, fringes
occupy distinct positions from photometric residues. Actually, for
high degree modes, a coupling between photometric noise and
oscillation signal still exists, since high degree spherical
harmonics extend largely in the Fourier domain. Therefore, part of
the energy in the largest spatial frequencies may be filtered
during this operation. Nevertheless, higher frequency mode
information will be still available, since most of its signal is
not cancelled by filtering operations, but may suffer from
amplitude estimate uncertainty.

In order to limit the damage generated by filtering the noise, the
spatial filter is as smooth as possible. It consists of an
ellipse, which major axis inclined of $-72^\circ$ with respect to
vertical, which includes only the photometric stains and the
central region. Besides, as for Jupiter, filter's boundary is
apodized to avoid rebounds by applying the inverse Fourier
transform, when getting back to the image plane. The resulting
couple of interference patterns, after spatial filtering is
presented in Fig (\ref{jupiter_U_V_brut}). Now, fringes are
centered around the 0 value. Hence, requirements for deconvoluting
the motionless phase are achieved.
\begin{figure}
\includegraphics[width=8.4cm]{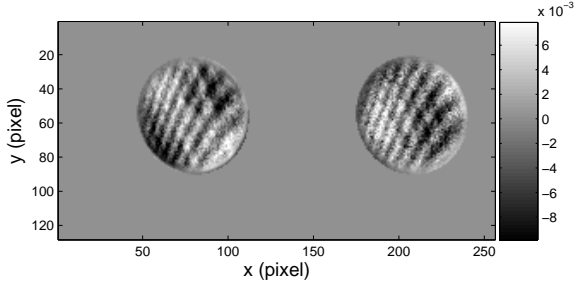}
\caption{Rough interferograms $(U\ind{jup},V\ind{jup})$ obtained
following Eqts. \ref{U} and \ref{V}. The interference pattern is
clearly visible, but some photometric residuals of same order of
magnitude as fringe's contrast, related to Jovian bands and zones,
still remain.} \label{jupiter_U_V_brut}
\end{figure}
\begin{figure}
\hskip .4 cm
\includegraphics[width=8cm]{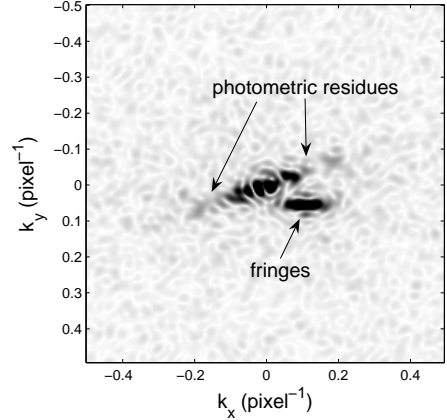}
\caption{Modulus of Fourier transform of the complex interferogram
$Z\ind{jup} = U\ind{jup} + \imath V\ind{jup}$. Photometric
residuals are located along the direction indicated by the two
upper arrows. This direction fits with Jupiter's inclination in
the detector's field. Main fringe information relies in the
horizontal line, under the central region.}
\label{jupiter_filtrage_fourier}
\end{figure}
\begin{figure}
\includegraphics[width=8.4cm]{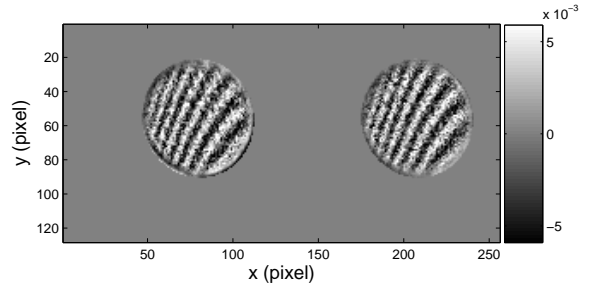}
\caption{The same couple of interferograms after filtering of the
photometric residuals. Note that fringes oscillate around 0, and
that contrast is about 0.6\%, which is very close to expected
values.  } \label{jupiter_U_V_filtre}
\end{figure}

\subsection{Jupiter velocity map}
The motionless fringe deconvolution is realized by the following
operation:
\begin{equation}
\label{deconvolutioninstr} Z\ind{j,rot} = Z\ind{jup}\times
Z_0^\ast
\end{equation}
The resulting interference pattern presents fringes parallel to
the planetary rotation axis (Fig. \ref{jupiterUVjrot}). Indeed,
the projection of Jupiter's velocity field $v = \Omega R(L)$,
where $\Omega$ is the angular velocity and $L$ the latitude,
towards the observer reduces to $v = \Omega x$, where $x$ is the
abscissa along Jupiter's equator. At zero order, Jovian rotation
can be considered as solid rotation since differential rotation
with respect to solid-body rotation is about 1\% on the equator.
Thus, $\Omega$ is almost uniform on the Jovian disk, and the phase
of the complex interferogram presents iso-values along the
rotation axis. Moreover, note that since noise level is about 900
ms$^{-1}$ per pixel, differential rotation is definitely invisible
on a single image.
\begin{figure}
\includegraphics[width=8.4cm]{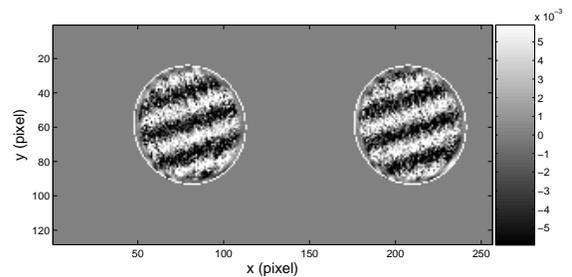}
\caption{Interference pattern after deconvolution of the
motionless component. Fringes present iso-values along the
rotation axis of Jupiter. The white circle indicates Jupiter's
size before resizing.} \label{jupiterUVjrot}
\end{figure}

Hence, a complex phaser which reproduces the solid rotation is
applied to each Jovian complex interferogram; it is defined by
$Z\ind{solid rot} = \exp(4\pi \sigma \Delta v\ind{rot}/c)$, where
$v\ind{rot} = 2\pi/T$ and where $T = 9$ h 55 m 30 s (system III)
is the mean rotation period. The main difficulty of such a process
lays on the accuracy of the estimate of Jupiter's position on the
detector, because 1 pixel corresponds to $350$ ms$^{-1}$. Two
methods have been envisioned to make the center of the solid
rotation phaser $Z\ind{solid rot}$ overlap on the center of
Jupiter's interferogram $Z\ind{j,rot}$. In both cases, a threshold
is applied to each image, in order to get rid of spurious
photometric signals, such as Jovian satellites of terrestrial
atmospheric light scattering. The ``barycenter'' method consists
in taking the coordinates of the barycenter of the total
photometric image $I = \sum_{i=1}^4 I_i$. The ``interspectrum"
method consists in determining the relative distance between two
images, taken at different dates, by measuring the phase $\Phi$ of
the interspectrum of the two images. Indeed, the phase of the
interspectrum of the couple of images $(I_1,I_2)$ is defined as:
\begin{eqnarray}
\Phi &=& \arg\left\{\mathcal F(I_1) \times \mathcal F(I_2)^\ast\right\}\\
 &\propto & (x_2-x_1) + (y_2-y_1)
\end{eqnarray}
where $(x_1,y_1)$ and $(x_2,y_2)$ are the coordinates of the
center of Jupiter and $\mathcal F$ and $\mathcal F^\ast$ indicate
the Fourier and inverse Fourier transformations. The second method
has been preferred because the barycenter estimate is too
sensitive to high spatial frequency photometric details, which
vary along the night, as cloud features or satellite transits. On
the contrary, the interspectrum method is sensitive only to low
spatial frequency.

After deconvolution of Jovian mean rotation (i.e. solid body
approximation), interferograms become flat since the remaining
phase $\phi = 4\pi\sigma \left(v\ind{J/S} + v\ind{E/J} + v\ind{E,
rot} + v\ind{osc}\right)$ is uniform across the field (Fig.
\ref{jupiterUVjplat}). The subtraction of $v\ind{J/S}$ and
$v\ind{E/J}$ is exposed in the next section, since they do not
change the noise to signal ratio of the phase map. Therefore, the
velocity map is retrieved thank to Eqt. 6 in paper I:
\begin{equation}
v = v_0 \arg\{Z\ind{flat}\} \label{magical_formula}
\end{equation}
with $v_0 \simeq 1$ km s$^{-1}$. Note that because of apodization
created by spatial filtering of photometric residuals, the entire
phase map is not exploitable. A part of the external region is cut
down, whose proportion is a function of the apodization strength.
Here, an external ring representing $1/4$ of the Jovian radius is
taken away (Fig. \ref{jupiterUVjplat}). As a consequence, the flux
is reduced by about of the partial elimination of Jupiter's is the
loss of 43 \% of the photons, which make the expected noise level
go up around 10.5 cm s$^{-1}$. However, the performance decrease
is limited by the low weight of external regions in the Doppler
signal.

The standard deviation of velocity across the Jovian disk is about
890 ms$^{-1}$ per pixel, that is to say 18.9 ms$^{-1}$ when
integrating the 2200 pixel of the resized Jovian disk. This
performance matches with expectations. If photon noise is reached,
a 7-hour night integration with 6-s sampling yields a noise level
as small as 29 cm s$^{-1}$.

\begin{figure}
\includegraphics[width=4.5cm]{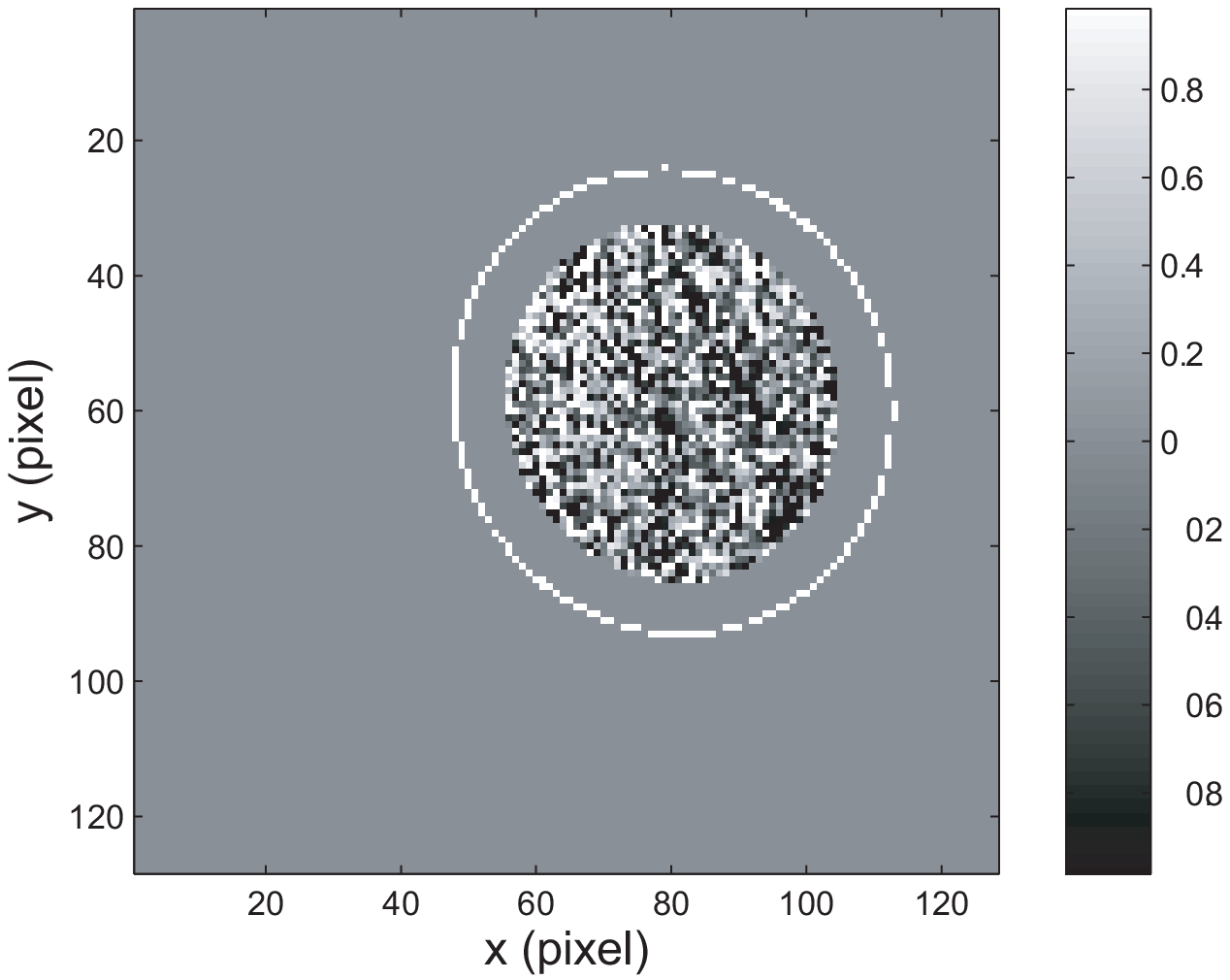}
\includegraphics[width=4.cm]{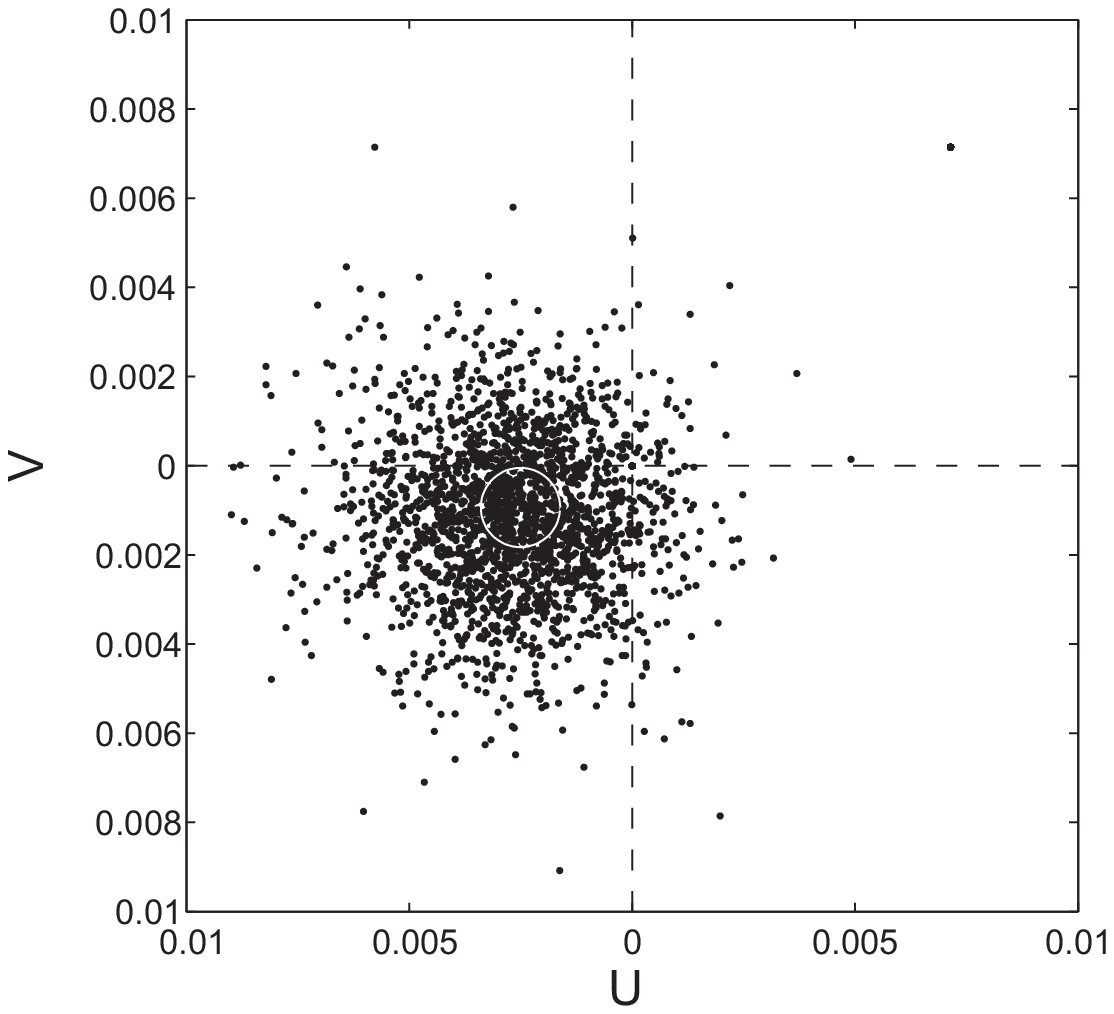}
\caption{Left: Jupiter phase map, after deconvolution of
motionless phase and Jovian rotation. Terrestrial motion and
temperature variations have not been taken into account, since
they only shift the whole phase by a uniform value. Right: complex
diagram of the same interferogram; $V\ind{flat}$ is plotted as a
function of $U\ind{flat}$. The white circle indicates the standard
deviation of the phase of the interferogram, which value is $8.9
10^{-4}$ rad. In terms of velocity, it corresponds 890 ms$^{-1}$
per pixel, that is to say 18.9 ms$^{-1}$ on the whole disk (2200
pixels are still available on Jupiter after resizing). }
\label{jupiterUVjplat}
\end{figure}

\section{Temporal analysis over one night}
\label{single_nite}

In the previous section, we have exposed the data processing
method for a single Jupiter quadruplet, for the extraction of the
Doppler signal. In this section, we analyse the time series of the
median phase extracted for each Jovian quadruplet. Then, we
identify the different sources of noise and present correction
methods.

\subsection{Phase global behavior}

The study of median phase along a single night allows us to
evaluate the mean noise level and to identify spurious signals.
Median phases are extracted from Jupiter's flat interferograms
$Z\ind{flat}$ as follow:
\begin{equation}
\phi\ind{med} = \arctan\left(\frac{V\ind{med}}{U\ind{med}}\right)
\end{equation}
where the subscript ``med" indicates the median value of the
considered variable. This estimate of Jupiter median phase has
been preferred to the direct median of the phase map, in order to
avoid noise coming from $2\pi$ jumps (see Fig.
\ref{jupiterUVjplat}). In Fig. \ref{phase_tf_one_night}, median
phase along night 9 and its power spectrum have been plotted.

First, according to data processing chain (Eqts. \ref{phase},
\ref{Z_flat_1} and \ref{Z_flat_2}), phase measurement is
proportional to the redshift of spectral lines, it appears that
the signal is overwhelmed by a strong low frequency noise. Indeed,
along the night, the velocity variation should be dominated by the
Earth rotation (409 m/s amplitude at Teide observatory), so the
phase shall increase of about 0.41 rad instead of decreasing of
about 1.12 rad. This implies that an unexpected stronger low
frequency drift dominates. Besides this low frequency noise, the
power spectrum highlights a rapid oscillation noise source around
6-mHz frequency.

Anyway, the standard deviation of median phase is around 27 m
s$^{-1}$ per image; images were taken every 6 s. By supposing an
only photon noise origin, the noise level reduces to 15 cm
s$^{-1}$for 7 nights of 7 hours, which is 1.3 time worse than
expected from previous section. In the following subsections, we
interpret the noise origins and describe the method which has been
used to get rid of their main effects.
\begin{figure}
\includegraphics[width=8.4cm]{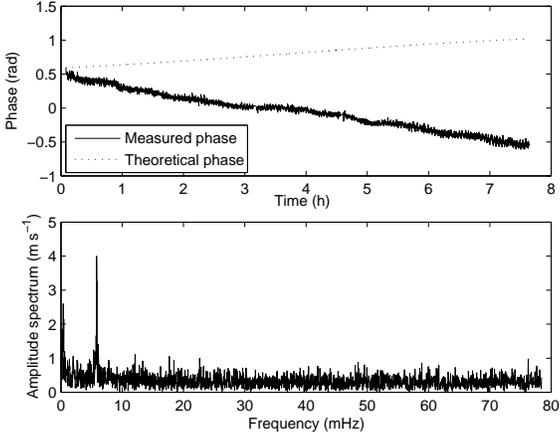}
\caption{Top. Median phase along night 9: measured phase (full
line) versus theoretical phase drift due to Earth rotation and
relative motion of Jupiter toward the Sun and the Earth (dashed
line). Bottom. Amplitude spectrum of the measured phase as a
function of the time frequency. Two spike forests blow out the
mean noise level at less than 0.5 mHz and at 6 mHz. Note that mean
noise becomes totally flat in the frequency range above 20 mHz,
which means that photon noise is achieved, around 41 cm s$^{-1}$.}
\label{phase_tf_one_night}
\end{figure}

\subsection{Temperature effect}

Paper I has expose that the interferometer is made of two pieces
of two different glasses, specially chosen as to compensate the
index variations and the dilatation, in order to have a stable
OPD. In particular, a temperature variation of about 1$^\circ$C
should have no effect on OPD in 14$^\circ$C environment.

The median of Jupiter's median phase over each night has been
plotted as a function of the corresponding median temperature
(Fig. \ref{temperature_phase_mediane}). Phase and temperature
appear clearly correlated. A variation of 1$^\circ$C introduces a
phase shift of 0.33 rad (i.e. 330 m s$^{-1}$), which is much above
than expected (between $-60$ m s$^{-1}$ and 30 m s$^{-1}$ for
temperature between 0 and 20$^\circ$C; see paper I). In fact, such
a thermal effect is retrieved when taking into account the error
bars upon dilatation coefficient and refractive index thermal
dependance ($\pm$ 10 \%).

Unfortunately, the correlation coefficient reported in Fig.
\ref{temperature_phase_mediane} does not allow us to correct
Jovian phase map from its thermal dependence. Indeed, the
comparison of phase and temperature over one night does not match
correctly: a several hour delay is still present (Fig.
\ref{temperature_phase_one_night}). This discrepancy is due to the
fact that temperature measurements do not correspond to the
Mach-Zehnder prism, but to the metallic box which holds it on.
Because of a greater thermal inertia, glass actual temperature is
time-shifted with respect to metal temperature. Therefore, a low
frequency filter is applied to data in order to reduce the noise
in the frequency range below 0.2 mHz.

Beyond a mean OPD drift, temperature variations generate
differential OPD variations with respect to light incidence angle
onto the prism. Since Jovian interferograms get deconvoluted from
motionless fringes with the help of a constant interferogram (Eqt.
\ref{deconvolutioninstr}), the differential OPD variations
introduce a slowly varying inclined surface in Jovian flat phase
maps (Fig. \ref{plan_incli}). This surface is bent along the
$x$-axis because OPD is only $x$-dependent. Such an inclined
surface yields a high frequency perturbation correlated to
Jupiter's position in the observed field. This noise is reduced by
fitting the spurious surface on each phase map, by a plane
inclined only with respect to $x$-axis. Thereafter, Jupiter's
phase map are set flat by using a smoothed estimate of the fitted
parameters.
\begin{figure}
\includegraphics[width=8.4cm]{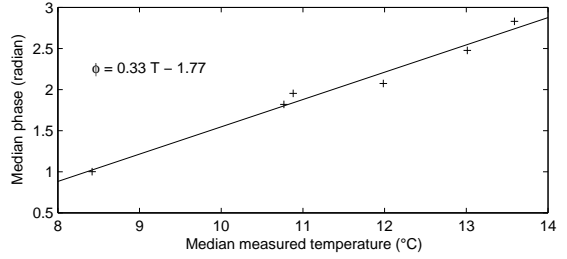}
\caption{Phase-temperature correlation. For each night, we have
plotted the median of the median phase as a function of the median
temperature over the entire night. The straight line represents
the least square estimate of the correlation between phase and
temperature.} \label{temperature_phase_mediane}
\end{figure}
\begin{figure}
\includegraphics[width=8.4cm]{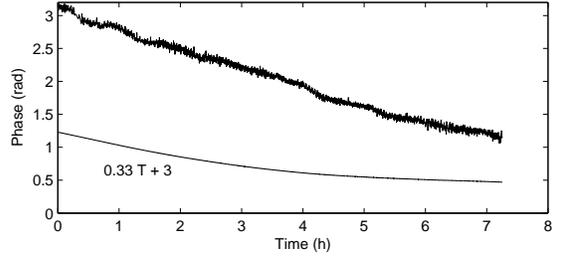}
\caption{Jovian median phase, corrected of Earth's rotation
component, over night 10 (full noisy line) and temperature
converted to phase with the previously fitted correlation
coefficient (full smooth line). Note that measured phase follows
the temperature with a several hour delay.}
\label{temperature_phase_one_night}
\end{figure}
\begin{figure}
\includegraphics[width=8.4cm]{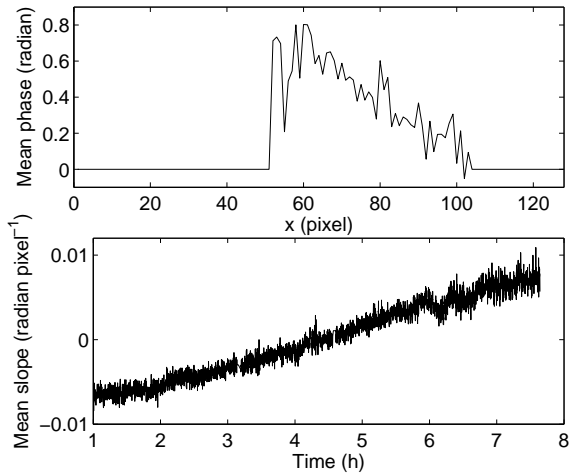}
\caption{Top: phase map averaged along y-axis as a function of the
x-axis. A negative slope is clearly visible. Bottom: root mean
square estimate of the slope of the inclined plane along night
10.} \label{plan_incli}
\end{figure}

\subsection{Guiding noise}
\begin{figure}
\includegraphics[width=8.4cm]{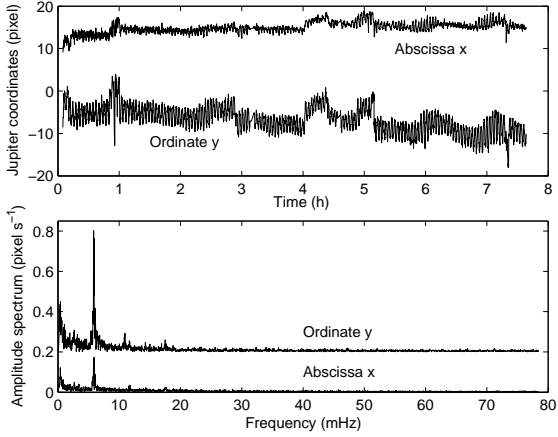}
\caption{Top. Jupiter's coordinates $(x,y)$ in the CCD field over
night 9. Note that $y$ varies much more than $x$ coordinate
because Jupiter is rotated about $-72^\circ$ on the image and
because most of the telescope guiding problems occur with the
right ascension. Bottom. Amplitude spectrum of both coordinates.
They present strong components below 6 mHz and significative
spikes up to 18 mHz.} \label{guiding_noise}
\end{figure}
The so-called ``guiding noise'' is the perturbation related to
Jupiter's position, which should not occur if Jupiter were at
fixed position. Theoretically, SYMPA's Doppler velocity
measurements are not sensitive to Jupiter's position onto the
field, because after motionless fringe deconvolution (Eqt.
\ref{deconvolutioninstr}) nothing should depend on the
coordinates. However, the comparison between spectra of Jupiter's
phase and coordinates $(x,y)$ (Figs. \ref{phase_tf_one_night} and
\ref{guiding_noise}) shows a strong correlation at 6 mHz. In fact,
many other correlated signatures occur at many frequencies.

This guiding noise has two main origins. First, the temperature
high frequency noise cannot be totally cancelled because of the
inclined surface fit inaccuracy. Second, Jupiter's position is
estimated around 1/15 of pixel. Therefore, the deconvolution of
Jovian rotation introduces a spurious signal correlated to Jovian
position. Thus, the guiding noise is a combination of these two
effects, for which it is hard to determine which of them
predominates.

The guiding noise is strongly reduced by removing the
intercorrelation of Jupiter's phase and coordinates. The
decorrelated phase with respect to both coordinated
$\phi\ind{decorr/x/y}$ is obtained as follow:
\begin{eqnarray}
\phi\ind{decorr/x}   &=& \mathcal F^{-1}\{(1-I\ind{x})\mathcal F\{\phi\}\} \\
\phi\ind{decorr/x/y} &=& \mathcal F^{-1}\{(1-I\ind{y})\mathcal
F\{\phi\ind{decorr/x}\}\}
\end{eqnarray}
where $\phi\ind{decorr/x}$ indicates the phase decorrelated with
respect to $x$, and $I\ind{x}$ and $I\ind{y}$ the interspectra
relative to $x$ and $y$, defined by:
\begin{eqnarray}
I\ind{x} &=&\frac{2}{\pi}\ \frac{\displaystyle
\arcsin\left(\Re\{\mathcal
F(\phi)\mathcal F^\ast(x)\}\right)}{|\mathcal F(\phi)||\mathcal F^\ast(x)|}\\
I\ind{y} &=&\frac{2}{\pi}\ \frac{\displaystyle
\arcsin\left(\Re\{\mathcal F(\phi\ind{decorr/x})\mathcal
F^\ast(x)\}\right)}{|\mathcal F(\phi\ind{decorr/x})||\mathcal
F^\ast(x)|}
\end{eqnarray}
Note that we apply a 75\% threshold upon $I\ind{x}$ and $I\ind{y}$
in order to decorrelate only significant guiding noise spikes.

Jupiter's phase standard deviation drops down from 27 m s$^{-1}$
per image to 21 m s$^{-1}$, which is very close to the expected
photon noise level (19 m s$^{-1}$ per image). Hence, by supposing
the remaining noise is only due to photons, the noise level
reduces to 32-cm s$^{-1}$ on a 7-hour integration and to 12 cm
s$^{-1}$ after 7 nights. The remaining not-photon noise is
ignored.

\section{The search for oscillations: no evidence for a Jovian signal}
\label{identificationmodes}

\subsection{Decomposition into spherical harmonic base}
A stationary oscillation mode can be described as the sum of two
spherical harmonics of degree $\ell$ and order $\pm m$. Jupiter
velocity field related to $p$-modes follows such a description.
Therefore, the Doppler signature of the radial velocity field
expands in a base made of the projected complex spherical
harmonics towards the observer.

The coefficients $c_\ell^m$ associated to each spherical harmonics
$Y_\ell^m$ are obtained as follow:
\begin{eqnarray}
c_\ell^m = \displaystyle \frac{\sum\ind{pixels}
\Re\{Y_\ell^m\}\times v}{\sum\ind{pixels} \Re\{Y_\ell^m\}}\ +\
\imath\ \frac{\sum\ind{pixels} \Im\{Y_\ell^m\}\times
v}{\sum\ind{pixels} \Im\{Y_\ell^m\}} \label{clm}
\end{eqnarray}
where $v$ indicates the velocity map (Eqt. \ref{magical_formula}).
The normalisation coefficient is done with respect to the actual
number of pixel after resizing. The $c_\ell^m$ coefficient modulus
are expressed in m s$^{-1}$. Oscillation search is performed in
the spectra of all spherical harmonics up to the degree $\ell =
25$.

\subsection{Regular temporal grid}

Velocity maps are extracted from mean flat interferograms,
averaged within 30-s intervals, but calculated every 15 s. Such a
process has two reasons. On one hand, the mean noise level inside
velocity maps drops down of a factor $\sqrt{5}$ since 30 s
contains 5 images, which limits strongly the 2$\pi$ jumps which
appear when applying Eqt. \ref{magical_formula}. On the other
hand, it allows to use the fast Fourier transform (FFT) to
calculate the power spectra. The procedure is of great interest
because the search for modes up to 25 means 625 spectra of 27000
point in the time series. The spacing of data every 15 s imposes a
cut-off frequency at 16.3 mHz, which is well beyond the expected
p-modes (less than 3.5 mHz from Mosser 1995). The averaging within
30-s intervals is done in order to avoid spectrum leakage when
calculating the Fourier transform.

\subsection{Power spectrum of modes $(\ell=0,m=0)$ and $(\ell=1,m=0)$}
\begin{figure*}
\hskip 1. cm
\includegraphics[width=16cm]{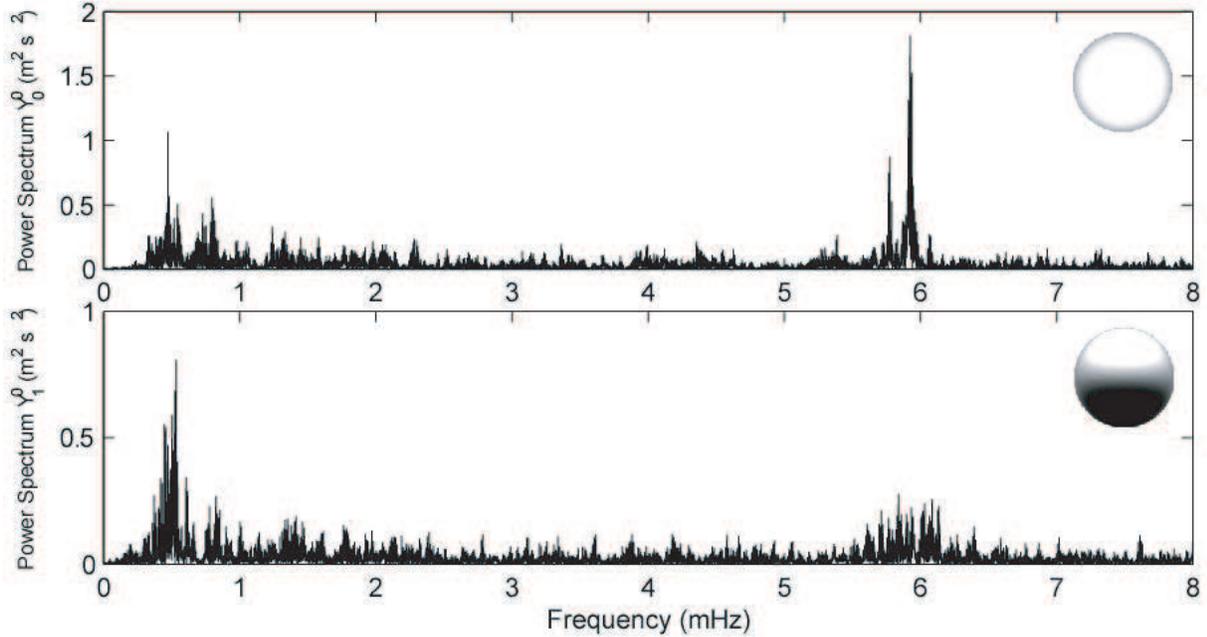}
\caption{Power spectrum of time series related to modes (0,0),
top, and (1,0), bottom, after concatenation of data along the
whole run. The mean noise level are respectively of about 12.6 cm
s$^{-1}$ and 11 cm s$^{-1}$, what matches with the last estimate
of the photon noise. The 6-mHz guiding spike still is about 1.3 m
s$^{-1}$ for the (0,0) mode, whereas it is only about 0.5 m
s$^{-1}$ for the (1,0) mode.} \label{SP}
\end{figure*}

We present the power spectrum of the time series corresponding to
spherical harmonics $Y_0^0$ and $Y_1^0$, for all the data of
Canaries observation campaign. The choice of these two modes among
625, permits to present the two main types of spectra. The first
is sensitive to the remaining guiding noise, whereas the second is
much less sensitive. Indeed, since guiding noise is mainly due to
right ascension control defects, the mode $(\ell=1, m=0)$
(hereafter $(1,0)$) is less sensitive to these problems since
north and south Jovian hemisphere compensate (Fig. \ref{SP}).

Both spectra exhibit a flat noise level in the frequency range
$[1, 8]$ mHz, excepted around 6 mHz, where a guiding signal
subsists. Beyond 8 mHz, the averaging over 30 s cuts-off the
signal. Guiding signature is reduced from 4 m s$^{-1}$ to 1.3 m
s$^{-1}$ after decorrelation processes. The mean noise level is
about 12.6 cm s$^{-1}$ for mode (0,0) and 11 cm s$^{-1}$ for mode
(1,0), which squares with the last-estimated photon noise level.
Note that such a performance has never been reached on Jupiter and
proves that the instrument and the data processing chain works
efficiently.

As regards the comparison to previous observations of S91 and M93
and M00, no excess of power is present between in the $[1,2]$ mHz
frequency range. Moreover, no large spacing $\nu_0$ is
highlighted, whose value is estimated around 150 $\mu$Hz, and
which was detected around $136$ and $143$ $\mu$Hz, respectively,
by S91 and M00. Such a dissension with previous observations will
be analyzed in detail in a future work.

As regards the excess of power in the frequency range $[0.3,0.6]$
mHz, no indication for a Jovian origin can be furnished at this
step of the data analysis. It could be a remaining low frequency
noise, related to temperature and position. A global analysis over
all the modes up to degree $\ell = 25$ is required to determine
its origin and to highlight global signature as Jovian rotation
frequency or $\nu_0$ frequency.

\subsection{Global analysis: no evidence for a Jovian signal}
\begin{figure*}
\hskip .5 cm
\includegraphics[width=17cm]{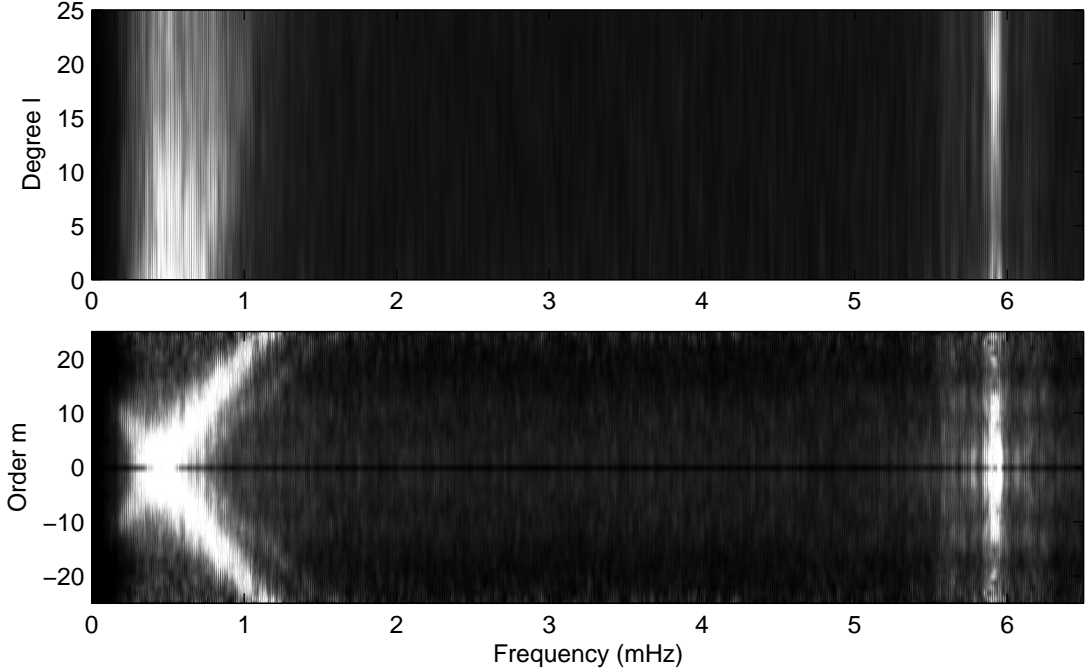}
\caption{Time-spatial frequency analysis. Top, the $(l, \nu)$
diagram represents the mean of all the power spectra at a given
degree $\ell$, as a function of all the explored degree. Bottom,
the $(m, \nu)$ diagram represents the mean of all power spectra at
a given order $m$, as a function of all explored orders. Note the
guiding noise is still present at 6 mHz, within a vertical line in
both graphics. The maximum amplitude is around 1.2 m s$^{-1}$.}
\label{diag}
\end{figure*}
\begin{figure}
\includegraphics[width=8.4cm]{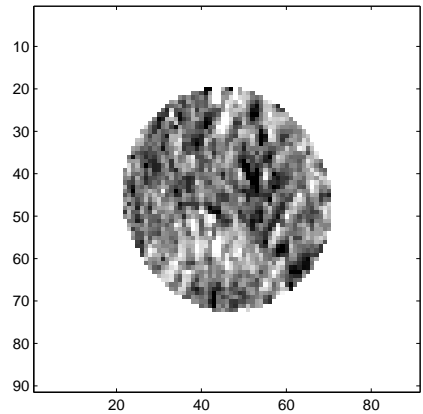}
\caption{Velocity maps averaged over 5 minutes. A fringe
structure, inclined of about Jupiter's inclination on the detector
is still present. The coupling of these feature with the spherical
harmonic masks may explain the observed features in the $(m,\nu)$
diagram. } \label{velocity_map_5min}
\end{figure}

As for helioseismology, simultaneous temporal and spatial
frequency analysis may reveal the presence of significant
information lost among noisy spectra. In Fig. \ref{diag} we
present the $(\ell, \nu)$ and $(m, \nu)$ diagrams. In the $(\ell,
\nu)$ diagram, the low frequency excess of power, picked out in
the $(0,0)$ and $(1,0)$ power spectra, is confirmed for almost all
degrees $\ell$ in the frequency range $[0.3, 1]$ mHz. It does not
exhibit an organized structure. Moreover, in comparison with past
observations, no signal is distinguishable in the $[1,2]$ mHz
range. On the other hand, in the $(m, \nu)$ diagram, the same
excess of power appears to be strongly structured. The energy is
distributed along 2 main lines, symmetric with respect to
abscissa, beginning at 0.5 mHz and ending at 1.2 mHz for $m=\pm
25$. Moreover, a second couple of lines, almost parallel to the
first ones, is still visible between frequency 0.7 and 1.6 mHz.
The mean slope of the principal lines is about 28 $\mu$Hz, which
corresponds to Jovian rotation frequency.

A quick analysis of the $(m, \nu)$ diagram gives some indications.
A simple guiding noise origin is excluded, because such a spurious
signal contaminates all the eigenmodes at the same frequency, as
for the 6 mHz spike (see Fig. \ref{diag}). Furthermore, guiding
noise becomes negligible beyond order $m = 15$, since its effect
compensates when applying high order spherical harmonic filtering
to velocity maps. At last, if the slope of about 28 $\mu$Hz could
be casual, it could indicate a Jovian origin to the observed
signal.

However, the Jovian signal hypothesis may vanish by supposing a
coupling between two spurious signals. First, it can be noticed,
after summation of velocity maps over 10-min, that a fringe
structure still remains, which periodicity fits almost with $m_0 =
16$ (Fig. \ref{velocity_map_5min}). Second, it can reasonably
supposed that photometric signal has not been totally removed
after the Fourier filtering step (see Sect. \ref{clean}). Thus, a
signal modulated by Jupiter's rotation probably underlies inside
velocity maps. The coupling of these two spurious signals
introduces a modulation of the signal by a $\cos( m_0 \Phi)$
factor, where $\Phi = \Omega\ind{r} t$ comes from the photometric
remnants; $\Omega\ind{r} = 2\pi\nu\ind{r}$ indicates the Jovian
rotation frequency. Consequently, when applying the spherical
harmonic search algorithm (Eqt. \ref{clm}), a coupling appears
between the $\cos( m_0 \Phi)$ modulation of velocity maps and the
$\cos( m \Phi)$ associated to $Y_\ell^m$. With this assessment,
velocity maps are modulated in the following way:
\begin{eqnarray}
v &=& v\ind{D}\ \cos( \Omega\ind{r}m_0 t)\cos( \Omega\ind{r}m t)\\
 &=& \frac{v\ind{D}}{2}\ \left\{\cos\left[ \Omega\ind{rot}(m_0 + m) t\right]+\cos\left[ \Omega\ind{rot}(m_0 - m)
 t\right]\right\}
\end{eqnarray}
Therefore, a linear dependence $\nu = m\ \nu\ind{r} + C$ appears,
where the constant term $C = m_0 \nu_r$ is about 450 $\mu$Hz,
which matches with the observed origin of the two lines.

\section{Conclusion and prospects}
\label{conclu}
\subsection{Instrumental performance}
The aim of SYMPA instrument was the detection and measurements of
acoustic modes on the giant planets of the solar system, with a
previously unequalled sensitivity around 4 cm s$^{-1}$. Such a
performance was estimated for a 16-day observation campaign with
50\% duty cycle (see paper I). By choosing to process only the
Canaries data, since the instrument used in San Pedro Martir
Observatory presents some defects which make the data more
difficult to process, the duty cycle is only 21\% over 10 nights.
In this case, the noise level is reevaluated at 10 cm s$^{-1}$. By
taking into account the lack of photons by a factor 2, underlined
in Sect. \ref{campagne}, the 1$-\sigma$ sensitivity decreases to
12 cm s$^{-1}$. After decorrelation of time series with respect to
Jupiter's position on the detector, outside the low frequency
range ($\leq 0.8$ mHz) and the 6-mHz spike, the power spectra are
flat and the mean noise level reaches 12 cm s$^{-1}$. Such a noise
level is 5 time better than previous observations which were
limited around 60 cm s$^{-1}$ (M00).

However, the data processing has highlighted some defects, as the
too strong optical path difference dependence to temperature and
some difficulties, as the separation between photometric and
spectrometric information. The latter make the extraction of
Doppler velocity hard to practise. This is mainly due to an
insufficient accuracy in rectifying the distortion, particularly
about the photometric effect of the distortion (variable PSF upon
the detector). This point suggests some instrumental
modifications, as the time modulation of OPD in order to modulate
the phase of the fringe pattern of about $\pi$. It would allow us
to replace the spatial subtraction between interferograms by a
time subtraction. Moreover, the spectral information would be
emphasized more easily by narrowing the entrance filter, which
would affect the global sensitivity. This would increase the
fringe contrast.

\subsection{Jupiter}
Jupiter's seismological observation of S91, M93 and M00 have all
exhibited an excess of power in the frequency range $[1,2]$ mHz,
which matches with the theoretical expectations (e. g. Mosser et
al. 1996). Moreover, they did not provide any spatial resolution.
If both observation methods (sodium cell and Fourier transform
spectrometry at fixed OPD) have presented an excess of power in
the same frequency range, the large spacing $\nu_0$ estimate was
very noisy and differed quite sensitively with respect to the
theoretical value of 153 $\mu$Hz (Gudkova et al. 1995).

Our observations do not present features close to $p$-modes
signature: the absence of the large spacing $\nu_0$ in power
spectra and $(\ell,\nu)$ diagram is significative. However, it is
worth to notice that because of the resizing of Jovian velocity
maps, the projected spherical harmonic base becomes a quite
degenerated base. Therefore, information from a given mode goes
diluted into other mode spectra, which may diminish strongly the
$p$-mode signature and the identification of specific related
structures. A way of leaving degeneracy has to be developed.

Moreover, Bercovici and Schubert (1987) roughly estimated
Jupiter's oscillation velocity amplitude between a few cm s$^{-1}$
and 1 m s$^{-1}$. Therefore it is not abnormal that oscillations
are not enlightened with a 12-cm s$^{-1}$ noise level, with a 21\%
duty cycle. This reduces the observed amplitude of  any
oscillations by a factor $(0.2)^{0.5}$. In these conditions, only
a 25-cm s$^{-1}$ signal could be detected at 1$-\sigma$ level.

\subsection{Prospects}
SYMPA has demonstrated to work properly, after taking into account
its technical defects (temperature dependance, field distortions,
difference of intensity between the two polarized outputs, lower
sensitivity of the CCD). It has permitted to reach the best remote
sensing velocity measurements upon giant planets. Some
improvements are to be performed on the existing instrument, as a
better thermal insulation and a slightly modified optical design
in order to reduce the geometrical distortions. An alternative and
more efficient solution is to rebuilt a prism, equipped with an
OPD time modulation.

From a most general point of view, this seismometer is a
tachometer, which furnishes velocity maps, instead of point to
point measures, which is the case of echelle-spectrometers. Thus,
it can be used to other kind of observations, such as wind
velocity measurements. SYMPA has tested an extra seismological
application in November 2007, by participating to a ground-based
observation campaign organised in sustain to ESA Venus-Express
probe, in order to characterize the lower mesosphere wind
velocity.

{\it What future for giant planet seismology} is the natural
question arising at the end of the first run of the first project
entirely dedicated to this topic. The main obstacles to such a
kind of measurements are the temporal coverage and the atmospheric
turbulences. Temporal coverage introduces a windowing effect which
makes the signal amplitude drop down and the deconvolution of
noise with respect to signal very hard because of the spreading of
information into large frequency ranges. Atmospheric turbulences
limit the spatial resolution and, above all, make the planet move
in the field, which generates guiding noise. A significant
improvement would arise with Antarctic observations: Schmider et
al. (2005) showed than a 80\% duty cycle can be reached for more
than 3 months. However, Jupiter south hemisphere oppositions will
last till 2009, and afterwards will not occur before 2018.

The ideal opportunity would come from space measurements aboard an
interplanetary spacecraft cruising towards Jupiter. With an
optical design similar to SYMPA, a 10-cm entrance pupil observing
for 2 months at a mean distance of 0.2 AU to Jupiter would allow
to reduce the noise level to few mm s$^{-1}$ and to increase the
spatial resolution till degree $\ell=100$. Such an instrumental
concept has been proposed to the European project of mission to
Europa and the Jupiter system, Laplace, which participates to the
ESA's Cosmic Vision programme (Blanc et al. 2006). The concept,
so-called ECHOES, is derived from the ground based instrument
SYMPA and the space helio-seismometer MDI onboard on SOHO space
Solar observatory (Scherrer et al. 1995). It would produce
simultaneously intensity maps of reflected solar light, velocity
maps and polarimetric maps of the whole visible surface with a
spatial resolution of about 500 km at the surface of Jupiter.
Also, such an instrument would be an efficient tool to measure the
wind velocity of the upper troposphere and the lower stratosphere.
Such a spatial project needs a feasibility study. On one hand,
from an instrumental point of view. Are the specific requirements
of seismology compatible with the interplanetary probe programme?
Is the envisioned method is the most adequate to space
measurements, or COROT-type photometric measurements (Baglin et
al. 1998) should be preferred as suggested by Mosser et al. (2004)
and Gaulme \& Mosser(2005) ? On the other hand, theoretical
improvements have to be performed about the modes excitation
mechanism, in order to evaluate the amplitudes of the expected
oscillations.


\end{document}